\documentclass{article}
\usepackage{amsmath,amsfonts,amsthm,color,amssymb,mathtools}
\usepackage{graphicx}
\usepackage{hyperref}

\theoremstyle{remark}
\newtheorem{remark}{Remark}
\usepackage[normalem]{ulem}
\newtheorem{example}{Example}
\usepackage[ruled, linesnumbered]{algorithm2e}
\usepackage{geometry}
\usepackage{enumerate}
\usepackage{bm}
\usepackage{subcaption}
\usepackage{rotating}
\usepackage{pdflscape}
\usepackage{tikz}
\usepackage{float}
\usetikzlibrary{matrix,shapes,chains,positioning,decorations.pathreplacing,arrows}
\usepackage{todonotes}
\usepackage{parskip}

\newcommand{\RR}{\mathbb{R}}

\DeclareMathOperator*{\argmin}{\arg\!\min}

\newcommand{\varphinn}{\varphi_\textrm{NN}}

\newcommand*\colvec[3][]{
	\begin{pmatrix}\ifx\relax#1\relax\else#1\\\fi#2\\#3\end{pmatrix}
}

\newcommand{\norm}[1]{\left\lVert#1\right\rVert}
\DeclareMathOperator{\diag}{diag}

\begin{document}
\title{On deep calibration of (rough) stochastic volatility models\thanks{The
    authors are grateful to Ben Wood, Jim Gatheral and Ryan McCrickerd for
    stimulating discussions. MT acknowledges financial support from the « Econophysique et Systèmes Complexes» chair under the aegis of the Fondation du Risque, a joint initiative by the Fondation de l’\'Ecole Polytechnique, l’\'Ecole Polytechnique and Capital Fund Management. CB and BS are grateful for financial support by the
   DFG through research grants BA5484/1 and FR2943/2. The present paper
   combines and consolidates findings of its two predecessor papers,
   \cite{BS18} and \cite{HMM19}.
}}

\renewcommand\footnotemark{}
\author{Christian Bayer\\
\large{TU Berlin and WIAS}\\
\normalsize{christian.bayer@wias-berlin.de}
\and Blanka Horvath\\
\large{Department of Mathematics, King's College London}
\\
\normalsize{blanka.horvath@kcl.ac.uk, b.horvath@imperial.ac.uk}
 \and Aitor Muguruza \\
\large{Department of Mathematics, Imperial College London
 \& NATIXIS}
\\
\normalsize{aitor.muguruza-gonzalez15@imperial.ac.uk}\\ 
\and Benjamin Stemper\\
\large{TU Berlin and WIAS}\\
\normalsize{benjamin.stemper@wias-berlin.de                                                   }\vspace{9pt}\\
 Mehdi Tomas\\
\large{CMAP \& LadHyx, \'Ecole Polytechnique}\\
\normalsize{mehdi.tomas@polytechnique.edu}
}

\maketitle


%

\begin{abstract}
  Techniques from deep learning play a more and more important role for the important task of calibration of financial models. The pioneering paper by Hernandez [Risk, 2017] was a catalyst for resurfacing interest in research in this area. In this paper we
  advocate an alternative (two-step) approach using deep learning techniques solely to learn the pricing map -- from
  model parameters to prices or implied volatilities -- rather than directly the calibrated model parameters as a function of observed market data.  Having a fast and accurate neural-network-based approximating pricing map (first step), we can then (second step) use traditional model calibration algorithms. In this work we showcase a direct comparison of different potential approaches to the learning stage and present algorithms that provide a sufficient accuracy for practical use.
  We provide a first neural network-based calibration method for rough volatility models for which calibration can be done on the fly.
 We demonstrate the method via a hands-on calibration engine on the rough Bergomi model, for which classical
  calibration techniques are difficult to apply due to the high cost of all known numerical pricing methods.  Furthermore, we display and compare different types of sampling and training methods and elaborate on their advantages under different objectives. As a further application we use the fast pricing method for a Bayesian analysis of the calibrated model.
\end{abstract}

\noindent \textbf{2010 }\textit{Mathematics Subject Classification}: 60G15, 60G22, 91G20, 91G60, 91B25


\tableofcontents
\newpage

\section{Introduction}

\begin{figure*}
  \centering
  \includegraphics[width=0.7\linewidth]{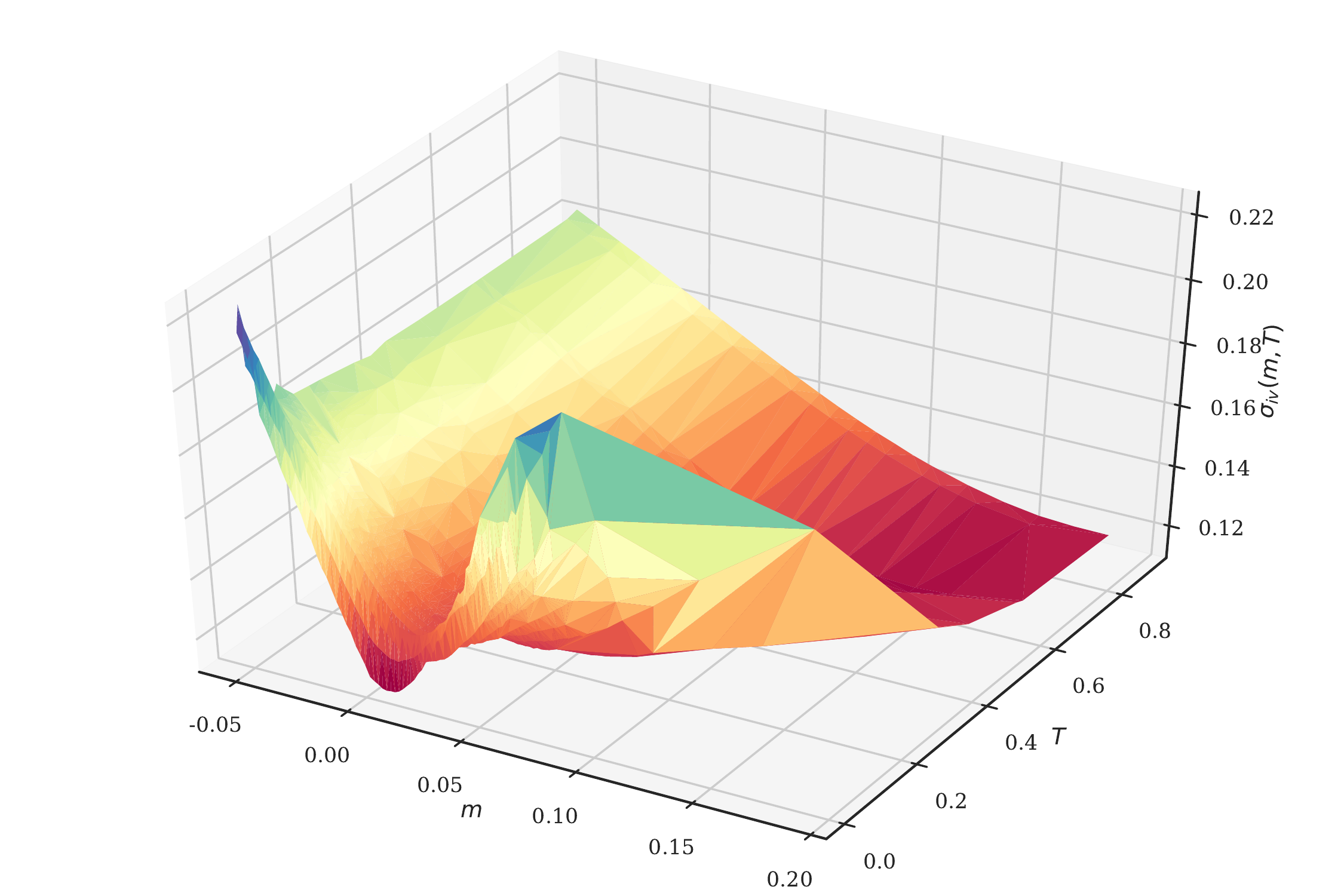}
  \caption{\textbf{SPX Market Implied Volatility surface on 15th
      February 2018.} IVs have been inverted from SPX Weekly European
    plain vanilla call mid prices and the interpolation is a
    (non-arbitrage-free) Delaunay triangulation. Axes denote
    log-moneyness $m = \log(K/S_0)$ for strike $K$ and spot $S_0$, time
    to maturity $T$ in years and market implied volatility
    $\sigma_\textrm{iv}(m,T)$.} 
  \label{fig:spxiv150218}
\end{figure*}

Almost half a century after its publication, the option pricing model by
Black, Scholes and Merton remains one of the most popular analytical
frameworks for pricing and hedging European options in financial markets. A
part of its success stems from the availability of explicit and hence
instantaneously computable closed formulas for both theoretical option prices
and option price sensitivities to input parameters (\emph{Greeks}), albeit at
the expense of assuming that \emph{volatility} -- the standard deviation of
log returns of the underlying asset price -- is deterministic and
constant. Still, in financial practice, the Black-Scholes model is often
considered a sophisticated transform between option prices and Black-Scholes
(BS) \emph{implied volatility (IV)} $\sigma_\textrm{iv}$ where the latter is
defined as the constant volatility input needed in the BS formula to match a
given (market) price. It is a well-known fact that in empirical IV surfaces
obtained by transforming market prices of European options to IVs, it can be
observed that IVs vary across moneyness and maturities, exhibiting well-known
smiles and at-the-money (ATM) skews and thereby contradicting the flat surface
predicted by Black-Scholes (Figure \ref{fig:spxiv150218}).  In particular,
Bayer, Friz, and Gatheral \cite{BFG15} report empirical at-the-money
volatility skews of the form
\begin{align}
  \label{eq: atmskewpower}
  \left|\frac{\partial}{\partial m}\sigma_\textrm{iv}(m,T)\right|  \sim
  T^{-0.4}, \quad T \to 0,
\end{align}
for log moneyness $m$ and time to maturity $T$.

While plain vanilla European call and put options often show enough liquidity
to be marked-to-market, pricing and hedging path-dependent options (so-called
\emph{exotics}) necessitates an option pricing model that prices European
options \emph{consistently} with respect to observed market IVs across
moneyness and maturities. In other words, it should parsimoniously capture
stylized facts of empirical IV surfaces. To address the shortcomings of
Black-Scholes and incorporate the stochastic nature of volatility itself,
popular bivariate diffusion models such as SABR \cite{Hagan1} or Heston
\cite{Hes93} have been developed to capture \emph{some}
important stylized facts. However, according to Gatheral \cite{Gat11}, diffusive
stochastic volatility models in general fail to recover the exploding
power-law nature \eqref{eq: atmskewpower} of the volatility skew as time to
maturity goes to 0 and instead predict a constant behaviour.

Sparked by the seminal work of \cite{AlosLeon, Fukasawa, GJR14}, we have
since seen a shift from classical diffusive modeling towards so-called rough
stochastic volatility models. They may be defined as a class of
\emph{continuous-path} stochastic volatility models where the instantaneous
volatility is driven by a stochastic process with H\"older regularity smaller
than Brownian Motion, typically modeled by a fractional Brownian Motion with
Hurst parameter $H<\frac{1}{2}$. The evidence for this paradigm
shift is by now overwhelming, both under the physical measure where time
series analysis suggests that log realized volatility has H\"older regularity
in the order of $\approx 0.1$ \cite{BLP16, GJR14} and also under the pricing
measure where the empirically observed power-law behaviour of the volatility
skew near zero may be reproduced in the model \cite{AlosLeon, BFG15, BFGHS, Fukasawa}. Serious computational and mathematical challenges arise
from the non-Markovianity of fractional Brownian motion, effectively forcing
researchers to resort to asymptotic expansions \cite{BFGHS, FZ17} in limiting regimes or (variance-reduced) Monte Carlo schemes \cite{BFGMS17, BFG15, HJM17, MP18} to compute fair option prices. This poses considerable bottlenecks for calibration of rough volatility models for practical purposes. One contribution of this work is to provide and explore different neural network based solutions to the task of fast calibration of rough volatility models.  

The solution we provide here is demonstrated on the rough Bergomi model but due to the nature of neural network approximations (as opposed to static polynomial approximations) it is fundamentally model agnostic and it consistently\footnote{By consistency we mean here that the proposed network (with the same architecture) can be trained on different models consistently without further modifications and yield satisfactory results irrespective of the chosen model for training. Our numerical experiments show that for the calibration of classical stochastic volatility models (SABR, Heston) a simpler network architecture is sufficient, while rough volatility models require a more nuanced network design.} carries over to other rough volatility models (of the same complexity) and to classical stochastic volatility models, which are by nature simpler to approximate.

The ``need for speed" is by no means limited to rough volatility models, although our initial motivation was indeed the rough Bergomi model. Parallel to this work, Ferguson and Green address in \cite[Section 1.1]{FG18} the ongoing struggle for faster pricing algorithms for more and more complex products and propose a deep learning approach to pricing basket options in a lognormal setting to achieve considerable speed-ups over Monte Carlo pricers. High dimensional problems as in \cite{FG18} are one useful applications of the speedup resuliting from this methodology. But it can also enable us to speed up more involved numerical methods for benchmark stochastic volatility models:  multiple integrals \cite{AKS19}, Monte Carlo-type methods \cite{RGLO17} or Finite Element Methods \cite{HR18} for the SABR model can thus compete in speed with the original SABR expansion formula \cite{Hagan1}, by pre-learning them through the DNN. 
In related contexts deep BSDE solvers have been used to replace Monte Carlo methods for solving Backward Stochastic Differential Equations in high dimension \cite{HJW17,PHL17,SSSz18} which can arise from pricing problem. Other authors used computational speedups provided by neural networks in the context of computationally expensive valuation adjustments \cite{Green15,PHL17}.
\\\\

The work we present here is very much in the spirit of the pioneering work of Avellaneda, Carelli and Stella \cite{ACS99}. Our focus in this work is on model calibration of stochastic volatiliy models and we propose computationally efficient and ready-to-use algorithms that can be applied to a variety of settings. Bearing in mind that deep neural network solutions are often challenged by concerns of generalisation and ``black-box-solutions", our goal is to limit the application of neural networks to parts of the calibration process that we can control and validate. As a first step, we identify the parts of the calibration process that are mainly responsible for the prevailing calibration bottlenecks, which we will replace by a deep neural network.
To motivate our approach, recall that model calibration is the optimization procedure of finding model parameters
such that the IV surface induced by the model best approximates a given market
IV surface in an appropriate metric. In the absence of an analytical solution,
it is standard practice to solve the arising weighted non-linear least squares
problem using iterative optimizers such as Levenberg-Marquardt (LM)
\cite{Levenberg, Marquardt}. However, these optimizers rely on the repetitive
evaluation of the function $\varphi$ from the space of model \& option
parameters (and external market information) to model BS implied
volatility. If each such evaluation involves a time-- and/or memory--intensive
operation such as a Monte Carlo simulation in the case of \emph{rough Bergomi}
\cite{BFG15} or other (rough) stochastic volatility models, this makes efficient calibration prohibitively expensive.

To bridge this computational gap and motivated by their prowess in approximating smooth functions \cite{UniversalApproxDerivatives}, neural networks have been used to build fast solutions to the calibration problem. 
Unsurprisingly, the tremendous rise in popularity of \emph{Deep learning} among academics and practitioners in recent years is closely tied to the widespread availability of cheap, high performance computing hardware as well as to theoretical advancements.
Fundamentally, most of the solutions in a calibration context build on the capability of multi-layered artificial neural networks to closely approximate functions $f$ only implicitly available through \emph{labeled datasets} of input-output pairs $\{(x_i, f(x_i))\}_{i=1}^N$.

In this context, we distinguish two kinds of approaches. The first, pioneered by Hernandez~\cite{Hernandez}, seeks to learn the mapping from implied volatility surfaces to model parameters (inverse problem) directly.  In ~\cite{Hernandez}, Hernandez proposes to use a neural  network to learn the complete calibration routine 
taking market data as inputs and returning calibrated model parameters, and calibrates the popular short rate model of Hull and White \cite{HW90} to market data in numerical experiments. In Section \ref{sec:hernandez} we describe this approach in more detail and perform a similar calibration experiment with the Heston Model.
In the rest of this paper, we will refer to it as the \textit{one-step} approach. 
In a second strand of research neural networks have been applied not directly to calibration problems, but simply to the obtain an approximative representation of derivative valuations, i.e. of option pricing maps: For example Hutchinson, Lo and Poggio \cite{Hutchison94} such as Culkin and Das \cite{CulcinDas17} applied neural networks to learn the Black-Scholes formula and McGhee demonstrates in \cite{McGhee} a neural networks representation of the lognormal SABR model.
In this paper we explore the advantages of shaping this second strand of research into a building block of a single \emph{two-step approach}.

The \textit{two-step} approach, which we highlight in this paper, first approximates the pricing map, (denoted, by $\varphi$ from model parameters to option prices) by a neural network (Step \textbf{(i)}) before calibrating the model, (via traditional calibration algorithms applied to the approximate pricing map $\varphinn$) to market data (Step \textbf{(ii)}).
Thereby we optimally leverage the capability of neural networks to approximate functions which are only implicitly available through input-output pairs $\{(x_i, \phi(x_i))\}_{i=1}^N$,
by training a fully-connected neural network on specifically tailored, synthetically generated training data to learn an approximative representation $\varphinn$ of the pricing functional $\varphi$. Details of this approach and its benefits are further explained in Section \ref{sec:pricing}.

There are different ways of approximating the pricing map (Step \textbf{(i)}): One could consider
a \textit{pointwise} approach where strikes and maturities are input
parameters of the pricing map along model parameters. Alternatively to this we also explore the advantages of proceeding here instead with a \textit{gridwise} approach, by first setting strikes and
maturities before learning the map from model parameters to the implied
volatility surface (with corresponding strikes and maturities).
In this work we showcase a direct comparison these approaches to the learning stage (Step \textbf{(i)}) and present algorithms that provide a sufficient accuracy for practical use, but are computationally efficient enough for daily practice on a large scale:

In particular, in Section \ref{sec:separation} we compare different network architectures and sampling methods according to different modelling objectives. Among these, the grid-based approach is particularly designed for applicability and efficiency in every day calibration practice. The novelty of our grid-based approach will allow us to tackle the calibration problem with a remarkably small neural network (3 layers 30 neurons), which to the best of our knowledge is the smallest network in the literature to successfully solve the calibration/pricing task. As opposed to the aforementioned works in the literature, we do not resort to GPU's or heavy computational resources, since the architecture of the problem easily permits to run the code on a standard CPU. This in turn, opens the door to its practical implementation in the financial industry  without the need to update current hardware systems.\\\\

The overall benefits of the \emph{two-step approach} are plentiful:
\begin{itemize}
\item   First, evaluations of $\varphinn$ amount to cheap and almost
 instantaneous forward runs of a pre-trained network. Second, automatic
 differentiation of $\varphinn$ with respect to the model parameters returns
 fast and accurate approximations of the Jacobians needed for the LM
 calibration routine. Used together, they allow for the efficient calibration
 of \emph{any} (rough) stochastic volatility model including \emph{rough Bergomi}.
 \item The \textit{two-step} approach also has overwhelming risk management benefits. Firstly, we can understand and interpret the output of our neural network and
therefore test the output as a function of model parameters against traditional numerical methods. (Indeed, the
output values correspond to option prices in the model under consideration.)
The second overwhelming advantage is that existing risk management libraries of models remain valid with minimal modification. The neural network is only used as a computational enhancement of models, and therefore, the knowledge and intuition gathered in many years of experience with traditional models remains useful.
  \item The training becomes more robust (with respect to generalisation errors on unseen data). Additionally, the trained network is independent from market data, and, in particular, from changing market environments. 
 \item  We can train the network to synthetic data -- model prices or implied
    volatilities computed by any adequate numerical method. In particular, we
    can easily provide as large training sets as desired.
 \end{itemize}


Both generating the synthetic data set as well as the actual neural network training are expensive in time and computing resource requirements, yet they only have to be performed a single time. Trained networks may then be quickly and efficiently saved, moved and deployed.  
We demonstrate this first advantage in a further application: a Bayesian calibration experiment, which is facilitated by our ability to nearly instantaneously call functional evaluations of option prices in a given model. To quantify the uncertainty about model parameter estimates obtained by calibrating with $\varphinn$, we infer model parameters in a Bayesian spirit from (i) a synthetically generated IV surface and (ii) SPX market IV data. In both experiments, a simple (weighted) Bayesian nonlinear regression returns a (joint) posterior distribution over model parameters that (1) correctly identifies sensible model parameter regions and (2) places its peak at or close to the true (in the case of the synthetic IV) or previously reported \cite{BFG15} (in the case of the SPX surface) model parameter values. Both experiments thus confirm the idea that $\varphinn$ is sufficiently accurate for calibration.

The paper is organised as follows: In Section~\ref{sec:an-abstract-view} we
present an abstract point of view on model calibration in finance. In
Section~\ref{sec:pricing} we give an overview of applications of techniques
from deep learning to model calibration. We also introduce our own framework
and discuss possible advantages and disadvantages as compared to other
approaches. In Section~\ref{sec:implementation} we focus on the concrete
implementation of our methods, both for the learning and for the actual
calibration stage. Numerical experiments are then presented in
Section~\ref{sec:Numerics}. In addition, we also apply the network in a
Bayesian approach. The Appendix~\ref{sec:Inverse Map} contains a numerical
comparison with an alternative deep learning approach to calibration.

\section{Model calibration}
\label{sec:an-abstract-view}

\emph{Calibration} describes the procedure of tuning model parameters to fit
a model surface to an empirical implied volatility surface obtained by
transforming liquid European option market prices to Black-Scholes implied
volatilities. A mathematically convenient approach consists of minimizing the
weighted squared differences between market and model implied volatlities of
$N \in \mathbb{N}$ plain vanilla European options.

Suppose that a model is parametrized by a set of parameters $\Theta$, i.e., by
$\theta \in \Theta$. We refer to Example~\ref{ex:rBergomi} for a concrete
example. Furthermore, we consider options parametrized by a parameter
$\zeta \in Z$. E.g., for put and call options we generally have
$\zeta = (T,k)$, the option's maturity and log-moneyness. There might be
further parameters which are needed to compute prices but can be observed on
the market and, hence, do not need to be calibrated. For instance, the spot
price of the underlying, the interest rate, or the forward variance curve in
Bergomi-type models (see \cite{BergomiBook}) falls under this type. For this
quick overview, we ignore this category. We introduce the \emph{pricing map}
\begin{equation*}
  (\theta, \zeta) \mapsto P(\theta, \zeta),
\end{equation*}
the price of an option with parameters $\zeta$ in the model with parameters
$\theta$. We are also given market prices $\mathcal{P}(\zeta)$ for options
parametrized by $\zeta$ for a (finite) subset $\zeta \in Z^\prime \subset Z$
of all possible option parameters. \emph{Calibration} now identifies a model
parameter $\theta$ which minimizes a chosen distance $\delta$ between model
prices $\left( P(\theta, \zeta) \right)_{\zeta \in Z^\prime}$ and market
prices $\left( \mathcal{P}(\zeta) \right)_{\zeta \in Z^\prime}$, i.e.,
\begin{equation*}
  \widehat{\theta} = \argmin_{\theta \in \Theta} \delta\left( \left( P(\theta,
      \zeta) \right)_{\zeta \in Z^\prime},  \left( \mathcal{P}(\zeta)
    \right)_{\zeta \in Z^\prime} \right).
\end{equation*}

\begin{remark}
  \label{rem:price=IV}
  Financial practice often prefers to work with implied volatilities rather
  than option prices, and we will also do so in the numerical parts of this
  paper. For the purpose of this introduction, any mentioning of a
  \emph{price} may be, mutatis mutandis, replaced by the corresponding implied
  volatility.
\end{remark}

In fact, the most usual way to choice of a distance function $\delta$ is a
suitably weighted least squares function, i.e.,
\begin{equation*}
  \widehat{\theta} = \argmin_{\theta \in \Theta} \sum_{\zeta \in Z^\prime}
  w_\zeta \left(P(\theta, \zeta) - \mathcal{P}(\zeta)\right)^2.
\end{equation*}
Here, the weights $w_\zeta$ can be chosen in order to reflect importance of an
option at $\zeta$ and the reliability of the market observation
$\mathcal{P}(\zeta)$. For instance, a reasonable choice might be the inverse
of the bid-ask spread (see \cite{Cont10} for a motivation), which puts low weight on prices of illiquid options.

As long as the number of model parameters is smaller than the number
$|Z^\prime|$of calibration instruments, the calibration problem is an example
of an overdetermined non-linear least squares problem, usually solved
numerically using iterative solvers such as the de-facto standard
Levenberg-Marquardt (LM) algorithm \cite{Levenberg,Marquardt}. Let $\bm{J} =
\bm{J}(\theta)$ denote the Jacobian of the map $\theta \mapsto \left(
  P(\theta, \zeta \right)_{\zeta \in Z^\prime}$ and let 
\begin{equation*}
  \bm{R}(\theta) \coloneqq \left( P(\theta, \zeta) - \mathcal{P}(\zeta)
  \right)_{\zeta \in Z^\prime}
\end{equation*}
denote the residual, then the Levenberg-Marquart algorithm iteratively
computes increments $\Delta \theta_k \coloneqq \theta_{k+1} - \theta_k$ by
solving  
\begin{equation}
  \label{eq:lev_mar}
  \left[\bm{J}(\mu_k)^T \bm{W}\bm{J}(\mu_k) + \lambda \bm{I}\right] \Delta \theta_k =
  \bm{J}(\mu_k)^T \bm{W}\bm{R}(\mu_k)
\end{equation}
 where $\bm{I}$ denotes the identity matrix, $\bm{W} =
 \diag\left(w_\zeta\right)$, and $\lambda \in \mathbb{R}$.

\begin{algorithm}[]
  \label{algo:levmar}
  \SetAlgoLined
  \KwIn{Implied vol map $\bm{\widetilde{P}}$ and its Jacobian
    $\bm{\widetilde{J}}$, market quotes $\bm{\mathcal{P}}$ }
  \SetKwInOut{Parameters}{Parameters}
  \Parameters{Lagrange multiplier $\lambda_0 > 0$, maximum number of
    iterations $n_\textrm{max}$, minimum tolerance of step norm
    $\varepsilon_\textrm{min}$, bounds $0 < \beta_0 < \beta_1 < 1$} 
  \KwResult{Calibrated model parameters $\theta^\star$} initialize model
  parameters $\theta = \theta_0$ and step counter $n=0$\; compute
  $\bm{\widetilde{R}}(\theta) = \bm{\widetilde{P}}(\theta) - \bm{\mathcal{P}}$
  and $\bm{\widetilde{J}}(\theta)$ and solve normal equations
  \eqref{eq:lev_mar} for $\Delta \theta$\; \While{$n < n_\textrm{max}$ and
    $\norm{\Delta \theta}_2 > \varepsilon$}{ compute relative improvement
    $c_\theta = \frac{\norm{\bm{\widetilde{R}}(\theta)}_2 -
      \norm{\bm{\widetilde{R}}(\theta +
        \Delta\theta)}_2}{\norm{\bm{\widetilde{R}}(\theta)}_2 -
      \norm{\bm{\widetilde{R}}(\theta) + \bm{\widetilde{J}}(\mu)
        \Delta\theta}_2}$
    with respect to predicted improvement under linear model\;
    \lIf{$c_\theta \leq \beta_0$}{ reject $\Delta\theta$, set
      $\lambda = 2\lambda$} \lIf{$c_\theta \geq \beta_1$}{ accept
      $\Delta\theta$, set $\theta = \theta + \Delta\theta$ and
      $\lambda = \frac{1}{2}\lambda$} compute $\bm{\widetilde{R}}(\theta)$ and
    $\bm{\widetilde{J}}(\theta)$ and solve normal equations \eqref{eq:lev_mar}
    for $\Delta\theta$\;
    
    set $n = n + 1$\;
  }
  \caption{Levenberg-Marquart calibration}
\end{algorithm}

It is hence necessary that the \emph{normal equations} \eqref{eq:lev_mar} be
quickly and accurately solved for the iterative step $\Delta \theta_k$. In a
general (rough) stochastic volatility setting this is problematic: The true
implied volatility map as well as its Jacobian $\bm{J}$ are unknown in
analytical form. In the absence of an analytical expression for
$\Delta \theta_k$, an immediate remedy is:
\begin{itemize}
\item[(I)] Replace the (theoretical) true pricing (or implied volatility) map
  $P$ by an efficient numerical approximation $\tilde{P}$ such as Monte Carlo,
  Fourier pricing. 	
\item[(II)] Apply finite-difference methods to $\tilde{P}$ to compute an
  approximate Jacobian $\tilde{\bm{J}}$. 
\end{itemize}
In particular, in many (rough) stochastic volatility models such as the
\emph{rough Bergomi model} (see Example~\ref{ex:rBergomi}), expensive Monte
Carlo simulations have to be used to approximate the pricing map. In a common
calibration scenario where the normal equations \eqref{eq:lev_mar} have to be
solved frequently, the approach outlined above thus renders calibration
prohibitively expensive.\\

\begin{remark}
  \label{rem:DNN-gradient}
  We note that many modern tensor-based machine learning frameworks are
  ideally suited for calibration tasks because the directly provide gradients
  of the output variable by use of automatic differentiation. 
\end{remark}


We would like to emphasize that our methodology can in principle be applied to
any model with finitely many parameters,
from the classical Black Scholes
or Heston models to the rough Bergomi model of \cite{BFG15}, also to large
class of rough volatility models (see Horvath, Jacquier and Muguruza
\cite{HJM17} for a general setup). In fact the methodology is not limited to
stochastic models, also parametric models of implied volatility could be used
for generating training samples of abstract models, but we have not pursued
this direction further.
For the sake of concreteness, we give an example of one rough volatility model, since
computational costs of available numerical methods are especially limiting for
this model class.

\begin{example}
  \label{ex:rBergomi}
  In the abstract model framework, the rough Bergomi model \cite{BFG15} is
  represented by
  $\mathcal{M}^{\mathrm{rBergomi}}(\Theta^{\mathrm{rBergomi}})$, with
  parameters $\theta = (\xi_0,\eta,\rho,H)\in \Theta^{\mathrm{rBergomi}}$. For
  instance, we may choose
  \begin{equation*}
    \Theta^{\mathrm{rBergomi}} = \RR_{>0} \times \RR_{>0} \times [-1,1] \times ]0,1/2[,
  \end{equation*}
  to stay in a truly rough setting. The model
  corresponds to the following system for the log price $X$ and the
  instantaneous variance $V$:
  \begin{subequations}
    \label{eq:rbergomi}
    \begin{align}
      dX_t&=-\frac{1}{2} V_t dt +\sqrt{V_t} dW_t,\quad \textrm{for} \ t>0, \quad X_0=0, \\
      V_t&=\xi_0(t)\mathcal{E}\left(\sqrt{2H}\eta \int_0^t
           (t-s)^{H-1/2}dZ_s\right),\quad \textrm{for} \ t>0, \quad V_0=v_0>0, 
    \end{align}
  \end{subequations}
  where $H$ denotes the Hurst parameter, $\eta>0$ ,
  $\mathcal{E}(\cdot)$ the Wick exponential, and
  $\xi_0(\cdot) >0$ denotes the initial forward variance curve (see
  \cite[Section 6]{BergomiBook}), and $W$ and $Z$ are correlated standard
  Brownian motions with correlation parameter $\rho\in [-1,1]$.
\end{example}

\section{Deep calibration}
\label{sec:pricing}

In the following sections we
elaborate on the objectives and advantages of this two step calibration
approach and present examples of neural network architectures, precise
numerical recipes and training procedures to apply the two step calibration
approach to a family of stochastic volatility models. We also present some
numerical experiments and report the learning errors compared to chosen parameters of the synthetic data.

There are several advantages of separating the tasks of pricing and
calibration. Above
all, the most appealing reason is that it allows us to build upon the
knowledge we have gained about the models in the past decades, which is of
crucial importance from a risk management perspective. By its very design,
\textbf{(i)} deep learning the \emph{price approximation} combined with
\textbf{(ii)} deterministic calibration does not cause more headache to risk
managers and regulators than the corresponding stochastic models do.
Designing the training as described above demonstrates how deep learning
techniques can successfully extend the toolbox of financial engineering,
without imposing the need for substantial changes in our risk management libraries.


\subsection{One-step approach: Deep calibration by the inverse map}
\label{sec:hernandez}
A more and more popular approach in quantitative finance (and many other
fields of engineering) is to develop purely data-driven frameworks, without
relying on formal models. This approach leaves the meaning of calibrated network
parameters unexplained, not to mention the ambiguity about the choice of the
number of network parameters and network design. This can cause major
challenges towards today's regulatory requirements. In addition, issues of
generalizability -- how can one price exotic options in a network trained with
vanilla option data, to give a simple example -- are difficult to analyse, and
traditional paradigms of finance -- such as no arbitrage -- are hard to
guarantee in the absence of a model.

A second, more model based approach was proposed in the pioneering work of
Hernandez \cite{Hernandez}, followed by several other authors such as
Stone~\cite{Stone}, Dimitroff, R\"oder and Fries~\cite{HestonConvolutional} and
many others. A main characteristic of the neural network proposed by
\cite{Hernandez} is that option price approximation and parameter calibration
are done in one step within the same network.
Indeed, the idea is to directly learn the whole calibration problem, i.e., to
learn the model parameters as a function of the market prices (typically
parametrized as implied volatilities). In the formulation of
Section~\ref{sec:an-abstract-view}, this means that we learn the mapping
\begin{equation*}
  \Pi^{-1}: \left( \mathcal{P}(\zeta) \right)_{\zeta \in Z^\prime} \mapsto \widehat{\theta}.
\end{equation*}
More precisely, \cite{Hernandez} trains a deep neural network based on
labelled data $(x_i, y_i)$, $i = 1, \ldots, N$, with
\begin{equation*}
  x_i = \left( \mathcal{P}(\zeta) \right)_{\zeta \in Z^\prime_i}
\end{equation*}
for day $t_i$ (in the past) and the corresponding labels
\begin{equation*}
  y_i = \widehat{\theta}_i,
\end{equation*}
obtained from calibrating the model to the market data $y_i$ using traditional
calibration routines. The number of labelled data points $N$ is, of course,
limited to the amount of (reliable) historical market price data available.

In spite of the promising results by Hernandez \cite{Hernandez} the main
drawback of this approach, as Hernandez observes, is the lack of control on
the function $\Pi^{-1}$. Furthermore, from a risk management
perspective one has no guarantee how well the learned mapping of
$\Pi^{-1}$ will solve the calibration problem when exposed to unseen
data. In fact, this is the behaviour observed in Hernandez \cite{Hernandez},
since the out of sample performance tends to differ from the in sample one,
suggesting a not fully satisfactory generalisation of the learned map. We
recover the same behaviour of the inverse map in our own experiments, which we
included in Appendix \ref{sec:Inverse Map}.

\subsection{Two-step approach: Learning the implied volatility map of models}
\label{sec:separation}

The two step approach is somewhere mid-way between a sole reliance on
traditional pricing methods (Monte Carlo, finite elements, finite differences,
Fourier methods, asymptotic methods etc.) and the direct approach described
above that calibrate directly to the price data. Here, one separates the
calibration procedure as described in Section~\ref{sec:an-abstract-view}
\textbf{(i)} We first learn (approximate) the pricing map by a neural network
that maps parameters of a stochastic model to prices or implied
volatilities. In other words, we set up and train (off-line) a neural network
to learn the pricing map $P$.  In a second step \textbf{(ii)} we calibrate
(on-line) the model -- as approximated by the neural network trained in step
\textbf{(i)} -- to market data using a standard calibration routine. To
formalise the two step approach, for an option parametrized by $\zeta$ and a
model $\mathcal{M}$ with parameters $\theta \in \Theta$ we write
$\widetilde{P}(\theta, \zeta) \approx P(\theta, \zeta)$ for the approximation
$\widetilde{P}$ of the true pricing map $P$ based on a neural network. Then,
in the second step, for a properly chosen distance function $\delta$ (and a
properly chosen optimization algorithm) we calibrate the model by computing
\begin{equation}\label{eq:ModelCalibration}
  \widehat{\theta} = \argmin_{\theta \in \Theta} \delta\left( \left(
      \widetilde{P}(\theta, \zeta) \right)_{\zeta \in Z^\prime}, \left(
      \mathcal{P(\zeta)} \right)_{\zeta \in Z^\prime} \right).
\end{equation}
In principle, this method is not unlike traditional calibration routines, as
the true option price has to be numerically approximated for all but the most
simple models. This particular approximation method tends to be orders of
magnitudes faster compared to other numerical approximation methods for all
tested models. In particular, note that the (slow) training stage of the
neural network itself only has to be done once. We will come back to
comparisons of actual computational times in the numerical section of this
paper. 

At this stage, we note that the deep calibration routine is not yet specified
in any details: apart from purely numerical details such as the choice of the
architecture of the neural networks, the loss functions and optimization
algorithms of both the training of the neural networks in stage \textbf{(i)}
and the actual calibration in stage \textbf{(ii)}, one particularly important
choice is whether the neural network learns implied volatilities of individual
options or rather a full implied volatility surface. Before discussing these
details, let us already highlight some of the differences to the one-step
approach of \cite{Hernandez}. While the one-step approach is probably marginally
faster, we see the main benefit of the two-step approach in the increased
stability, which is influenced by two key differences:
\begin{itemize}
\item As the neural network is only responsible for option pricing in the
  model, synthetic data can (and should) be used for training. Hence, we can
  easily increase the number of training data, and the training data are
  completely unpolluted from market imperfections.
\item The two-step approach induces a natural decomposition of the overall
  calibration error into a pricing error (from the neural network) and a model
  misfit to the market data. Hence, the performance of the neural network
  itself is generally independent of changing market regimes -- which might,
  of course, change the suitability of the model under consideration.
\end{itemize}
These points, in particular, imply that frequent re-training of the neural
network is not needed in the two-step approach.

\subsubsection{The two step approach: Pointwise training and implicit and grid-based training}

The underlying principle of the two-step approach appears in one way or another in a number of related contributions 
De Spiegeleer, Madan, Reyners and Schoutens~\cite{MadanSchoutens} and McGhee \cite{McGhee}. In fact, the early works of Hutchinson, Lo and Poggio \cite{Hutchison94} and the more recent work of Culkin and Das \cite{CulcinDas17}--where Deep Neural Networks are applied neural to learn the Black-Scholes formula--can be recognised as Step \textbf{(i)} of the two-step approach in a Black-Scholes context.
Also Ferguson and Green \cite{FG18} examine Step \textbf{(i)} of the two-step approach in \cite{FG18} for basket options in a lognormal context and observe that the network even has a smoothing effect and increased accuracy in comparison to the underlying Monte Carlo prices. In this section, we examine its advantages and present an analysis of the objective function with the goal to enhance learning performance. Within this framework, the pointwise approach has the ability to asses the quality of $\widetilde{P}$ using Monte Carlo or PDE methods, and indeed it is superior training in terms of robustness.
\vspace*{0.5cm}\\ 
\textbf{Pointwise learning}\\

\begin{enumerate}
\item[Step (i):] Learn the map
  $\widetilde{P}(\theta,T,k)=\widetilde{\sigma}^{\mathcal{M}(\theta)}(T,k)$ -- that is in equation \eqref{eq:ModelCalibration} above we have $\zeta = (T,k)$. In the case of vanilla options ($\zeta = (T,k)$) one can rephrase this learning objective as an implied volatility problem:
 In the implied volatility problem the more informative implied volatility map $\widetilde{\sigma}^{\mathcal{M}(\theta)}(T,k)$  is learned, rather than call- or put option prices $\widetilde{P}(\theta,T,k)$. We denote the
  artificial neural network by $F(w;\theta, \zeta)$ as a function of the
  weights $w$ of the neural network, the model parameters $\theta$ and the
  option parameters $\zeta$. The optimisation problem to solve is the
  following:
  \begin{equation}
    \label{eq:general_loss}
    \widehat{\omega}:= \argmin_{w\in\mathbb{R}^n}\sum_{i=1}^{N_{\mathrm{Train}}}
    \eta_i(\widetilde{F}(w;\theta_{i},T_i,k_i)-\widetilde{\sigma}^{\mathcal{M}}(\theta_{i},T_i,k_i))^2. 
  \end{equation}
  where $\eta_i\in\mathbb{R}_{>0}$ is a weight vector.
\item[Step (ii):] Solve the classical model calibration problem 
$$\widehat{\theta} \coloneqq \argmin_{\theta\in\Theta}
  \sum_{j=1}^{m} \beta_j(\widetilde{\sigma}^{\mathcal{M}(\theta)}(\theta,T_j,k_j)-\sigma^{\mathrm{MKT}}_{\mathrm{BS}}(k_j,T_j))^2.$$
\end{enumerate}
for some user specified weights $\beta_j\in\mathbb{R}_{>0}$, where now the (numerical approximation of the) option price $\widetilde{P}(\theta,T,k)$  resp. implied volatility $\widetilde{\sigma}^{\mathcal{M}(\theta)}(T,k)$ is replaced by the DNN approximation $\widetilde{F}(\widehat{\omega}; \theta, T, k)$ obtained in Step \textbf{(i)}.

The critical part is, of course, the first step, as the second one merely
corresponds to classical calibration against liquid options. For the first
step, key issues are the choice of training data and the architecture of the
neural network. Regarding the training data, the general idea is as follows:
\begin{enumerate}
\item Choose realistic \emph{``prior''} distributions for both model
  parameters $\theta$ and option parameters $\zeta$ ($= (T,k)$ in the above
  notation). The point is that many theoretically possible parameters are very
  unlikely to ever occur in real markets, for both model and option
  parameters. Hence, it is wasteful to spend resources to learn the pricing
  map for, say, maturities in the range of hundreds of years. The simplest
  choice is to simply impose uniform distributions on truncated parameter
  ranges, but nothing prevents more ``informed'' possibilities, for instance
  taking into account historical distributions of estimated model parameter
  values or observed option parameter values.
\item Simulate model and option parameters according to the distribution
  chosen before and compute the corresponding option price or implied
  volatility, which serves as label for the respective parameter vector. The
  computation can be done for any available numerical method, for instance
  Monte Carlo simulation. As an aside, this mechanism can, of course, be used
  to produce training, testing and validation data in the sense of the machine
  learning literature.
\end{enumerate}

\begin{remark}
  Note that the above mentioned ``informed'' parameter distributions could
  also be encoded as weights into the loss function for the training of the
  neural network.
\end{remark}

\begin{remark}
  Instead of simulation of parameter values, we could also consider
  deterministic grids in the parameter space. In very high dimensional
  parameter spaces this probably becomes unfeasible due to the curse of
  dimensionality, but in the current context this approach may very well
  improve training of the neural network. We leave a comparison to future
  work.
\end{remark}

\vspace*{0.5cm}
\textbf{Implicit \& grid-based learning}


We take this idea further and design an implicit form of the pricing map that
is based on storing the implied volatility surface as an image given by a grid
of ``pixels''. This image-based representation has a formative contribution in
the performance of the network we present in Section \ref{sec:Numerics}. 
Let us denote by
$\Delta:=\{k_i,T_j\}_{i=1,\;j=1}^{n,\;\;\;\;m}$ a fixed grid of strikes and
maturities, then we propose the following two step approach:
\begin{enumerate}
\item[Step (i):] Learn the map
  $\widetilde{F}(\theta)=\{\sigma^{\mathcal{M}(\theta)}_{BS}(T_i,k_j)\}_{i=1,\;j=1}^{n,\;\;\;\;m}$
  via neural network where the input is a parameter combination
  $\theta\in\Theta$ of the stochastic model $\mathcal{M}(\theta)$ and the
  output is a $n\times m$ grid on the implied volatility surface
  $\{\sigma^{\mathcal{M}(\theta)}_{\mathrm{BS}}(T_i,k_j)\}_{i=1,\;j=1}^{n,\;\;\;\;m}$
  where $n,m\in\mathbb{N}$ are chosen appropriately (see Section
  \ref{sec:architecture}) on a predefined fixed grid of maturities and
  strikes. $\widetilde{F}$ takes values in $\mathbb{R}^L$  where
  $L=\textrm{strikes}\times \textrm{maturities} = nm$.
  The optimisation problem in the image-based implicit learning
  approach is:
  \begin{equation}\label{eq:gridbased_loss}
    \widehat{\omega}:=
    \argmin_{w\in\mathbb{R}^n}\sum_{i=1}^{N_{\mathrm{Train}}^{\mathrm{reduced}}}
    \sum_{j=1}^{L} \eta_j(\widetilde{F}(\theta_{i})_j-\sigma^{\mathcal{M}}(\theta_{i},T_j,k_j))^2, 
  \end{equation}
where 
$N_{\mathrm{Train}}=N_{\mathrm{Train}}^{\mathrm{reduced}} \times L$ and $\eta_i\in\mathbb{R}_{>0}$ is a weight vector.  
\item[Step (ii):] Solve the minimisation problem $$\widehat{\theta}:=\argmin_{\theta\in\Theta}\sum_{i=1}^L \beta_j(\widetilde{F}(\theta)_{i}-\sigma^{\mathrm{MKT}}_{\mathrm{BS}}(T_i,k_i))^2,$$ 
for some user specified weights $\beta_j\in\mathbb{R}_{>0}$
\end{enumerate}


The data generation stage for the image-based approach works as in the
point-wise approach, except that the option parameters $\zeta = (T,k)$ are, fixed and are no longer part of the learning algorithms -- except
implicitly in the output/labels of the neural network. This is why they appear in the general objective function of pointwise learning \eqref{eq:general_loss} but no longer appear in the objective function \eqref{eq:gridbased_loss} of the grid-based learning  above.
In practice, we choose a grid $\Delta$ of size $8\times 11$. By evaluating the implied volatility surface along $8\times 11$ gridpoints with $40.000$ different parameter combinations in $\Theta$ we effectively evaluate the ``fit" of the surface as a whole. In our experiments we chose a $8\times 11$ grid for practical reasons, but we are by no means limited to this number. For example, to obtain even higher accuracy, one could also choose a coarser grid, which would require longer learning time, but recall that learning only has to be done once. One advantage of a grid-based sampling is that one can re-use the same set of generated Monte Carlo paths along grid points. Once a grid is fixed one can also easily refine the grid by adding further refined points to it using the same set of Monte Carlo paths (evaluated at more time points).

Clearly, the neural network does depend on the grid $\Delta$ of option
parameters $\zeta$. Hence, we need to interpolate between gridpoints in order to be able to calibrate (in the calibration Step \textbf{(ii)}) also to such options, whose
maturity and strike do not exactly lie on the grid $\Delta$. 
While in the pointwise training the \textbf{interpolation between sampling points} is done by the network $\widetilde{F}(\theta)$ automatically (both in the model parameter space $\Theta$ and along the implied volatility surface in $\mathbb{K}\times \mathbb{T}$), in the grid-based implicit learning the network is only used for interpolation in the parameter space $\Theta$, and it is implicit in the space dimension, that is, --based on smoothness assumptions of the implied volatility surface-- we
interpolate between gridpoints of the implied-volatility surface manually, using appropriate splines. This indirect dependence of the trained network on $\Delta$ is alluded to by the name ``implicit learning''.
\vspace*{0.2cm}

\textbf{Implicit smile-based learning: \\
--And outlook towards an implicit learning with more elaborate grids and tessalations of the IV surface--}\\ We note that McGhee \cite{McGhee} follows an \emph{implicit} approach  for the lognormal SABR model, which lies somewhere between the
pointwise and the image-based approaches of Step \textbf{(i)}: There, the inputs
are $(\theta^{\mathrm{SABR}}, T, k_1, \ldots, k_{10})$, and there are ten volatility
outputs ${\sigma_1, \ldots, \sigma_{10}}$ per maturity $T$. 
Since between the reference points of the smile McGhee \cite{McGhee} also interpolates (by splines) based on a smoothness assumption of implied volatilities, we also refer to this approach as \emph{implicit} training.
The reference points $k_1, \ldots, k_{10}$ on the volatility surface are determined as a direct functional of the model parameters $\theta^{\mathrm{SABR}}$ and of the maturity $T$, that is the learning is done slice-by slice. 
This sampling technique showcases an excellent working example of a \emph{representative functional sampling} on the surface, where more samples are taken in certain regions of the surface, to ensure a good accuracy of the training in those regions (e.g. regions with higher liquidity). Though the sampling of the strikes in \cite{McGhee} is bespoke to the SABR model, it motivates the idea of \emph{representative sampling grid (or tessalation net)}, which would be desirable to achieve also in a model agnostic context. We note that the introduction of the weight vectors $\eta_i \in \mathbb{R}_{>0}$ in the objective function \eqref{eq:gridbased_loss} of the grid-wise approach has a similar effect as a higher sampling frequency of a neighbourhood/point.

\subsubsection{The role of the objective function: Pointwise training versus implicit and grid-based training}

Comparing the pointwise approach (characterised by the general objective
function \eqref{eq:general_loss}) and the image-based approach (characterised
by the objective function \eqref{eq:gridbased_loss}), we find that
both of them can be advantageous in certain situations. We highlight the connection between the two below, and
elaborate on some of the respective advantages of each approach.\\

Equation \eqref{eq:gridbased_loss} can be brought to the form of
\eqref{eq:general_loss} equation by inserting (into \eqref{eq:general_loss})
the specification values $\theta = \theta^\prime$, with
$$
\theta^\prime_1 = \theta_1, \ldots, \theta^\prime_L = \theta_1,
\theta^\prime_{L+1} = \theta_2, \ldots,
$$
and recalling that
$L=\textrm{strikes}\times \textrm{maturities}$ and
$N_{\mathrm{Train}}=N_{\mathrm{Train}}^{\mathrm{reduced}} \times L$. Hence,
the pointwise approach is more general than the image-based one.

With this in mind we make the general note, many of the various advantages and disadvantages of both
approaches can, in principle, be mitigated by careful choice of the data
generation mechanism (of the training and validation datasets) and the loss
function in the training.

\begin{itemize}
\item The biggest difference, between pointwise and image based implicit learning procedures
is  that image based implicit learning requires an outside (implicit) interpolation
between the learned implied volatilities in order to compute the implied
volatility of an option with an arbitrary strike or maturity, not aligned
with the grid. At face value, this is of course an advantage of the
pointwise (explicit) approach, where the interpolation is rather performed by the deep
neural network. On the other hand, we note that the function
$(T,k) \mapsto \sigma^{\mathcal{M}}(\theta; T,k)$ (for fixed model
parameters $\theta$) is usually a very well understood smooth function. (At
least for useful models, as the market implied vol surface arguably is nice
and smooth.) This is not necessarily true for
$\theta \mapsto \sigma^{\mathcal{M}}(\theta; T,k)$, which is not nearly as
well understood for more modern sophisticated models such as rough Bergomi. Hence, we
have much more confidence in applying standard interpolation in $(T,k)$
rather than in $\theta$, which also lives in a higher dimensional
space. Hence, the outside interpolation may, in practice, not cause any
difficulties. 

\item Indeed, this very same structure induces a reduction of variance in the
training data for the image-based approach as compared to the pointwise
approach. Formally speaking, in the image based approach only the
model parameters are sampled, while the strike and maturities of the
underlying instruments are deterministic. As a side note, keep in mind that we
should always compare the two approaches based on a fixed number $N_{\mathrm{Train}}$
of total training data.

\item It is also easier to take into account the structure of real financial data
into the data generation for the pointwise approach by adjusting the (random) sampling distribution on the surface accordingly. Clearly, not all options
are equally \emph{important} for the purpose of calibration, but we would like
to concentrate on liquid options. It is easy to adjust the sampling
distribution for strikes and maturities in the pointwise approach to take into
account historical numbers of liquidity. In the grid-based approach, this can to some extent be taken into account by the choice of the weight vector $\eta_i \in \mathbb{R}_{>0}$ in \eqref{eq:gridbased_loss}, or more accurately taken into account by using non-uniform, non-tensorized, or bespoke quasirandom sampling
grids, with higher density of points in regions with higher liquidity. 


\item The image-based approach may be seen as an efficient dimension-reduction
technique as compared to the pointwise one. Indeed, as dimensions are shifted
from the input of the neural network to the output, the learning task becomes
easier since lower-dimensional. Of course, the price we pay is that we only
learn the values of the implied volatilities on a fix grid $\Delta$ of option
parameters. In this example, this price is, however, worth paying since the
regularity of the volatility surface is well understood. This implies that we
know very well the number and location of grid points required to get good
fits globally in terms of the chosen interpolation.
\end{itemize}

In the particular calibration example presented in Section \ref{sec:Numerics} below, the image-based
approach performed somewhat better than the pointwise approach, which indicates
that the variance and dimension reduction features may be more important than
the other aspects in the above comparison.

\begin{remark}
  \label{rem:chebyshev}
  In principle, the two-step approach is also amenable to other numerical
  interpolation methods. For instance, we could also use Chebyshev
  interpolation to approximate model implied volatilities such as \cite{Glau19}.
\end{remark}
\begin{remark}
  \label{rem:chebyshev-2}
  In line with Remark~\ref{rem:chebyshev}, we note that the image-based
  approach (in conjunction with the outside interpolation) is a hybrid between
  a pure DNN approximation such as the point-wise approach and a standard
  polynomial interpolation method, such as Chebyshev approximation, see
  \cite{Glau19} for example. Of course, other, more specialized interpolation
  methods on the implied volatility surface are also possible, for instance
  using the SVI volatility parameterization, see for example \cite{Itkin14}.
\end{remark}

\section{Practical implementation}
\label{sec:implementation}
We start by describing the approximation network (Step \textbf{(i)} of Section \ref{sec:pricing}  with objective functions \eqref{eq:general_loss} and \eqref{eq:gridbased_loss}) and leave the discussion of calibration (Step \textbf{(ii)}) for Section \ref{sec:calibrationStep} below.
While several realted works \cite{Hutchison94, CulcinDas17, McGhee} have demonstrated that learning the pricing map (Step \textbf{(i)}) in the Black-Scholes model and in certain clasical stochastic volatility models (such as the lognormal SABR model in \cite{McGhee}) can be done to a satisfactory accuracy with a single hidden layer, the situation is--as often--more delicate in the case of rough volatility models. Since these models are highly nonlinear nature, they also require deeper networks for an accurate approximation of their pricing functional.

\subsection{Network architecture and training}

\label{sec:architecture}



We present the architecture used for the grid-based approach in some detail,
as this approach was used for most of the numerical examples below.
\begin{enumerate}
\item A fully connected feed forward neural network with 3 hidden layers
  and $30$ nodes on each layers;
\item Input dimension = $n$, number of  model parameters
\item Output dimension = 11 strikes$\times$ 8 maturities for this experiment, but this choice of grid can be enriched or modified.
\item The three inner layers have $30$ nodes each, which adding the
  corresponding biases results on a number $$(n+1)\times30+ 3\times
  (1+30)\times 30+(30+1)\times88=30 n+5548$$ of network parameters to
  calibrate.
\item We choose the Elu $\sigma_{\mathrm{Elu}}=\alpha(e^x-1)$ activation
  function for the network.
\end{enumerate}


We train the neural network using gradient descent, the so-called `Adam'
minibatch training scheme due to Kingman and Ba \cite{KBAdam}, which is a
version of the Stochastic Gradient Descent algorithm. In the following, $w$ denotes the set of parameters --
weights and biases -- of a neural network $F = F(w,x)$.  Given parameters
$0 \leq \beta_{1}, \beta_{2} < 1, \epsilon$, $\alpha$, initial iterates
$u_{0} := 0, v_{0} := 0, w_{0}\in\Omega$, the Adam scheme has the following
iterates:
\begin{align*}
  g_{n} &  := \nabla^{w} \sum_{i = 1}^{m}
          \mathcal{L}\left(F(w_{n-1},X_{n,m}^{\text{batch}}),F^*(X_{n,m}^{\text{batch}})\right)\\ 
  u_{n+1} & := \beta_{1} u_{n} + (1 - \beta_{1})g_{n} \\
  v_{n+1} & := \beta_{2} v_{n} + (1 - \beta_{2}) g_{n}^{2} \\
  w_{n+1} & := w_{n} - \alpha \dfrac{u_{n+1}}{1 - \beta_{1}^{n+1}} \dfrac{1}{\sqrt{v_{n} / (1 - \beta_{2}^{n+1})} + \epsilon}.
\end{align*}

\subsection{The calibration step}
\label{sec:calibrationStep}

Once the pricing map approximator $\widetilde{F}$  for the implied volatility
is found, only the calibration step is left to solve. We use the
Levenberg-Marquart algorithm as presented in Section~\ref{sec:an-abstract-view}.


\subsubsection{Bayesian Analysis of the Calibration}
\label{sec:bayes-analys-calibr}


Intuitively, we are interested in \emph{quantifying the uncertainty} about
model parameter estimates obtained by calibrating with the approximative
implied volatility map map $\widetilde{F}$. To this end, we switch to a
Bayesian viewpoint and treat model parameters $\theta$ as random
variables. The fundamental idea behind Bayesian parameter inference is to
update prior beliefs $p(\theta)$ with the likelihood $p(\bm{y} \mid \theta)$
of observing a given point cloud $\bm{y}\in \mathbb{R}^N$ of implied
volatility data to deduce a posterior (joint) distribution
$p(\theta \mid \bm{y})$ over model parameters $\theta$.

Formally, for pairs $\left(T_i, k_i\right)$ of time to maturity and
log-moneyness, let an implied volatility point cloud to calibrate against be
given by
\begin{equation*}
  \bm{y} = \left[y_1\left(T_1, k_1\right), \ldots, y_N\left(T_N,
      k_N\right)\right]^T \in \mathbb{R}^N 
\end{equation*}
and analogously, collect model implied volatilities for model parameters
$\theta$
\begin{equation*}
  \bm{\widetilde{F}}\left(\theta\right) = \left[\widetilde{F}\left(\theta,
      T_1, k_1\right), \ldots, \widetilde{F}\left(\theta, T_N,
      k_N\right)\right]^T \in \mathbb{R}^N.
\end{equation*}
We perform a liquidity-weighted nonlinear Bayes regression. Mathematically,
for heteroskedastic sample errors $\sigma_i>0, i=1, \ldots, N$, we postulate  
\begin{equation*}
  \bm{y} = \bm{\widetilde{F}}\left(\theta\right) + \bm{\varepsilon}, \quad
  \bm{\varepsilon} \sim \mathcal{N}\left(0, \diag(\sigma_1^2, \ldots,
    \sigma_N^2) \right),
\end{equation*}
so that for some diagonal weight matrix $\bm{W} = \diag(w_1, \ldots, w_N) \in
\mathbb{R}^{N \times N}$, the liquidity-weighted residuals are distributed as
follows 
\begin{equation*}
  \bm{W}^{\frac{1}{2}}\left[\bm{y} -
    \bm{\widetilde{F}}\left(\theta\right)\right] \sim  \mathcal{N}\left(0,
    \diag(w_1\sigma_1^2, \ldots, w_N\sigma_N^2)\right).
\end{equation*} 
In other words, we assume that the joint likelihood
$p\left( \bm{y} \mid \theta \right)$ of observing data $\bm{y}$ is given by a
multivariate normal. In absence of an analytical expression for the posterior
(joint) probability $p(\theta|\bm{y}) \propto p(\bm{y}|\theta) p(\theta)$, we
approximate it numerically using MCMC techniques \cite{FHLG13} and plot the
one- and two-dimensional projections of the four-dimensional posterior by
means of an MCMC plotting library \cite{For16}.

\begin{remark}
  \label{rem:sum-squares-gaussian}
  Of course, from a statistical point of view, loss functions of sum of
  squares form corresponds to a normality assumption on the error distribution
  when interpreted as an MLE, for instance. The normality assumption above,
  hence, merely mirrors the common choice of sum-of-squares as loss function
  for calibration in finance.
\end{remark}

\section{Numerical experiments}
\label{sec:Numerics}

\subsection{Speed and accuracy of the price approximation networks}

As mentioned in Section \ref{sec:separation} one crucial improvement win in
comparison with direct neural network approaches, as pioneered by Hernandez
\cite{Hernandez}, is the separation of (i) the implied volatility
approximation function, mapping from parameters of the stochastic volatility
model to the implied volatility surface--thereby bypassing the need for
expensive Monte-Carlo simulations---and (ii) the calibration procedure, which
(after this separation) becomes a simple deterministic optimisation problem.

Table \ref{Table:NNSpeed} shows the CPU computation time for functional
evaluation of a full surface under the rough Bergomi model of
Example~\ref{ex:rBergomi}. Here, we take the forward variance $\xi_0$
as constant. In a future work we take a similar approach to constract a network that can consistently approximate a variaty of models including the the rough Bergomi model with a forward variance curve that is approximated (more generally) by piecewise constant function.

\begin{table}[!htpb]
  \centering
  \begin{tabular}{c|c|c|c}
    \multicolumn{1}{c|}{\begin{tabular}[c]{@{}c@{}}MC Pricing\\Full
                          Surface \end{tabular}} & 
    \multicolumn{1}{|c|}{\begin{tabular}[c]{@{}c@{}}NN Pricing\\  Full
                           Surface \end{tabular}} & 
    \multicolumn{1}{|c|}{\begin{tabular}[c]{@{}c@{}}NN Gradient\\ Full
                           Surface \end{tabular}} & 
    \multicolumn{1}{|c}{\begin{tabular}[c]{@{}c@{}}Speed up\\ NN
                          vs. MC \end{tabular}}  \\ 
    \hline 
    $500.000\;\mu \text{s}$ & $14,3\;\mu \text{s}$ & $47\;\mu\text{s}$ &
    $21.000-35.000$\\
  \end{tabular}
  \caption{Computational time of pricing map (entire implied volatility
    surface) and gradients via Neural Network approximation and Monte Carlo
    (MC) for the image-based approach}
\label{Table:NNSpeed}
\end{table}


Table~\ref{Table:NNSpeed} provides the speed of evaluating the trained neural
network for the image-based approach, the numbers for the pointwise approach
are very similar. We used
\begin{itemize}
\item Total number of parameteres: $5.668$
\item Training set of size $34.000$ and testing set of size $6.000$
\item Rough Bergomi sample: $(\xi_0,\nu,\rho,H)\in
  \mathcal{U}[0.01,0.16]\times \mathcal{U}[0.5,4.0]\times
  \mathcal{U}[-0.95,-0.1]\times \mathcal{U}[0.025,0.5]$ 
\item Strikes: $\{0.5,0.6,0.7,0.8,0.9,1,1.1,1.2,1.3,1.4,1.5\}$
\item Maturities: $\{0.1,0.3,0.6,0.9,1.2,1.5,1.8,2.0 \}$
\item Training data samples of Input-Output pairs are computed using Algorithm
  3.5 in Horvath, Jacquier and Muguruza \cite{HJM17} with $60.000$ sample
  paths and the spot martingale condition i.e. $\mathbb{E}[S_t]=S_0,\;t\geq 0$
  as control variate. 
\end{itemize}

\begin{figure}[!htpb]
\includegraphics[scale=0.45]{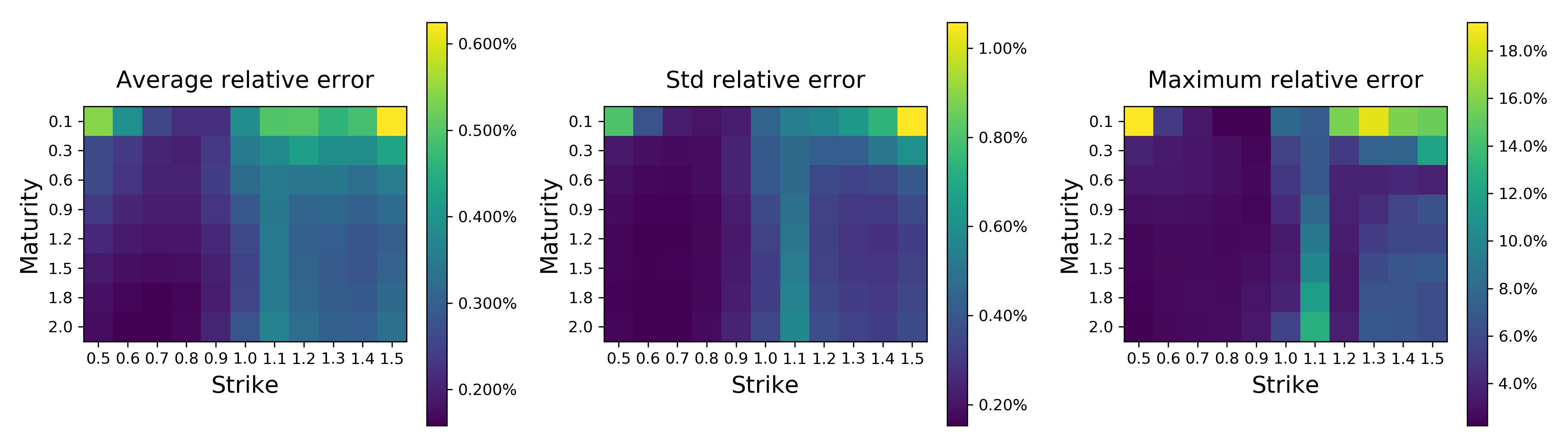}
\caption{We compare surface relative errors of the neural network approximator
  against the Monte Carlo benchmark across all training data ($34.000$ random
  parameter combinations)in the rough Bergomi model. Relative errors are given
  in terms of Average-Standard Deviation-Maximum (Left-Middle-Right).} 
\label{fig:rBergomi1}
\end{figure}

\begin{figure}
\begin{center}\includegraphics[scale=0.5]{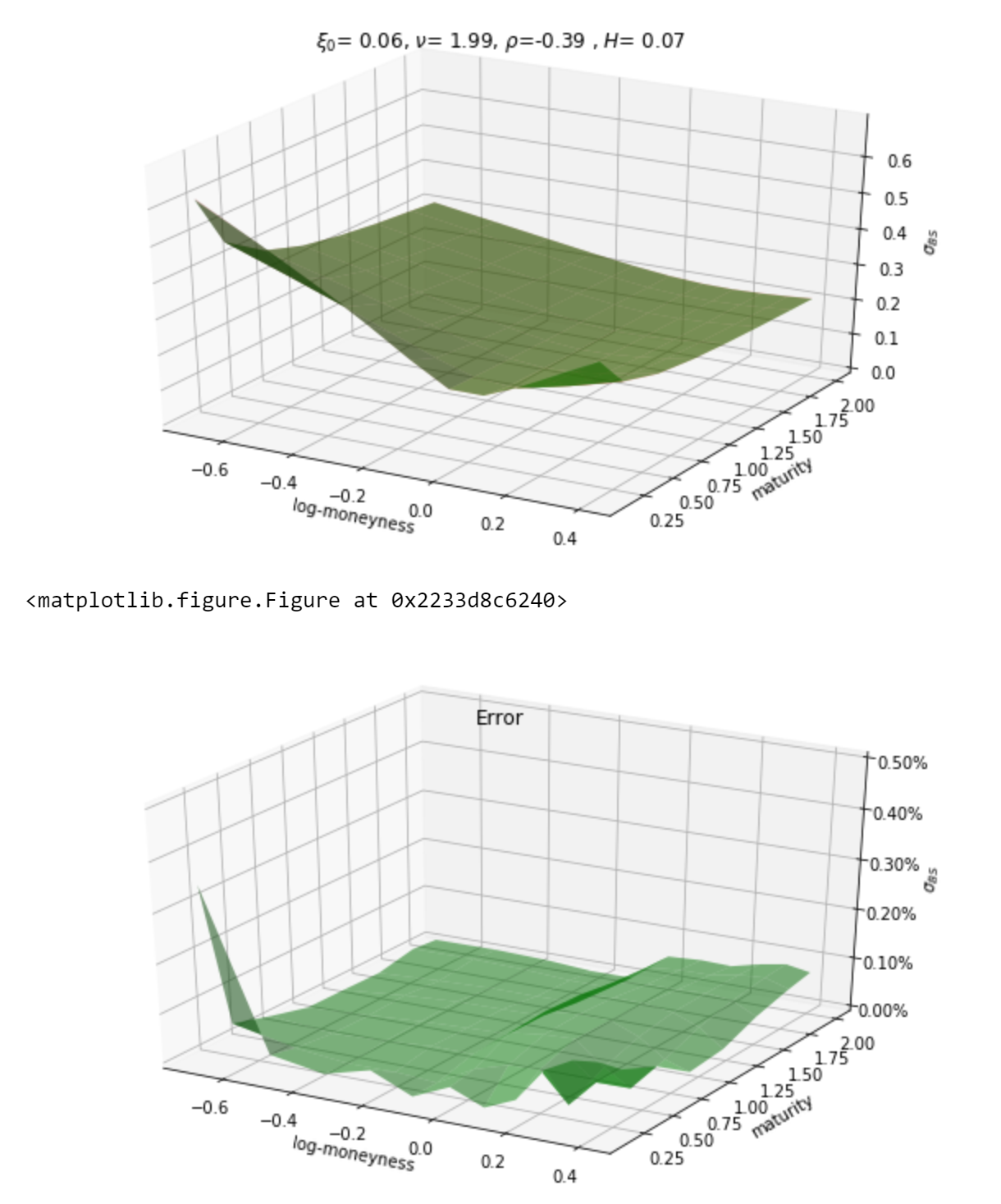}
\includegraphics[scale=0.5]{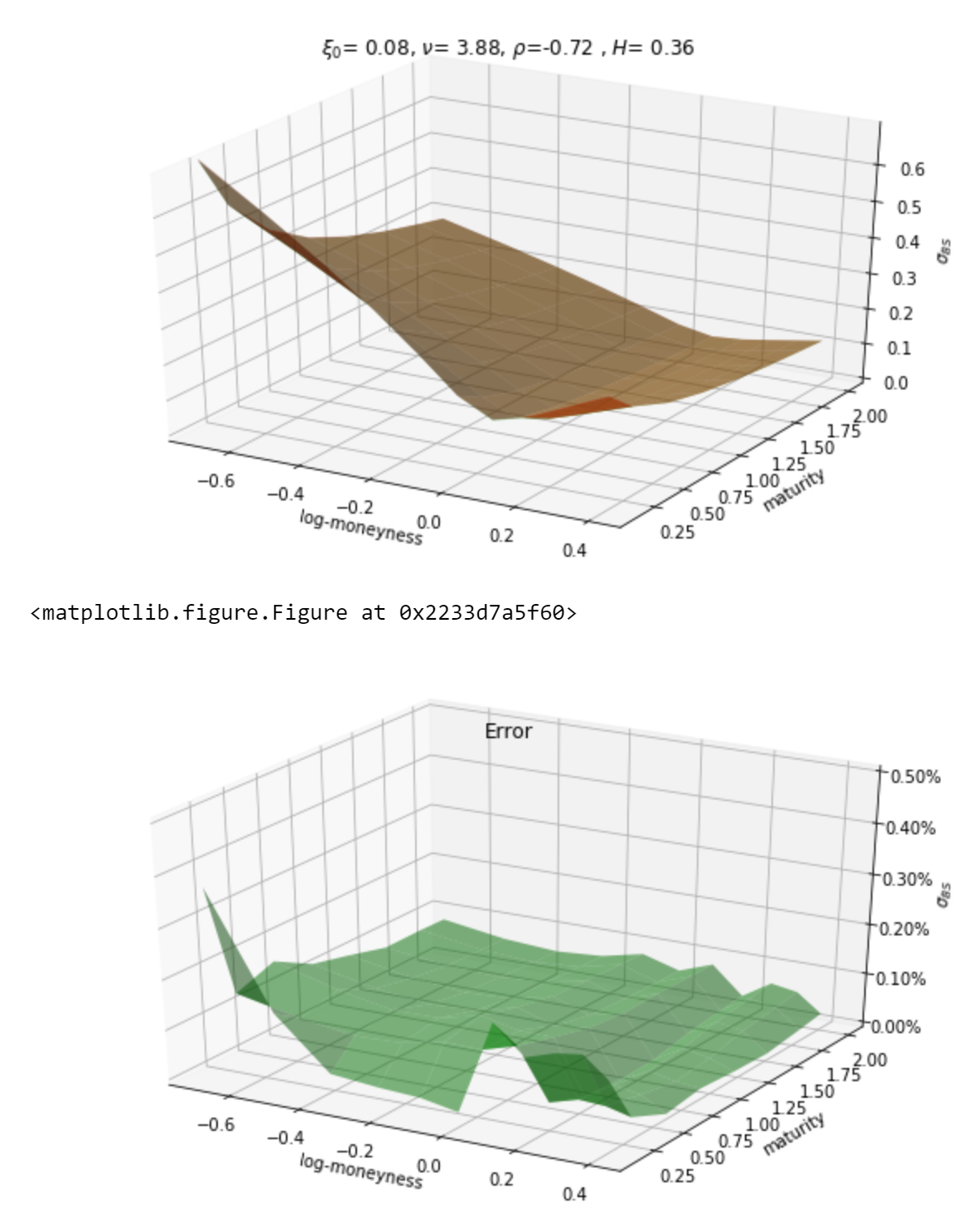}\label{Fig:surfaceerrors}
\end{center}
\caption{The Figure illustrates the distribution of the approximation error in space after the interpolation to a full implied volatility surface in two examples of model parameter choice.}
\end{figure}

Figure~\ref{fig:rBergomi1} show that the average (across all parameter
combinations) relative error\footnote{Relative here is computed here as
  $|\sigma^{NN}(T,k)-\sigma^{MC}(T,k)|/|\sigma^{MC}(T,k)|$.} between neural
network and Monte Carlo approximations is far less than $0.5\%$ consistently
(left image in Figure~\ref{fig:rBergomi1}) with a standard deviation of less
than $1\%$ (middle image in Figure~\ref{fig:rBergomi1}). Nevertheless, the
maximum relative error goes as far as $25\%$. As previously stated, the beauty
of this approach is the ability to asses whether the approximation is suitable
and if not, where exactly fails or is more delicate. In this case, we observe
that the approximation is less precise for short maturities and deep
out-of-the-money/in-the-money options. Theses errors are consistent with the
errors of the Monte Carlo training set 
.

\subsection{Calibration speed and accuracy}
\label{sec:calibr-speed-accur}
To demonstrate the advantage of our two-step approach we obtain calibration times less than 40 milliseconds for the full implied volatility surface in the rough Bergomi model, which was notoriously slow to calibrate (several seconds) by Monte Carlo methods due to its non-Markovian nature. Note that these calibration times become much lower (usually under 10 milliseconds) for Markovian stochastic volaility models. This considerable speedup is due to the 21000-35000 factor speedup (reported in Tabe 1) of the approximation network.

In order to asses calibration the accuracy compared to synthetic data in a
controlled experiment, the accuarcy of calibrated model parameters
$\widehat{\theta}$ compared to the synthetically generated data with the set
of parameters $\overline{\theta}$ that was chosen for the generation of our
synthetic data. We measure the accuracy of the calibration via parameter
relative error i.e.
\begin{equation*}
  E_R(\widehat{\theta})=
  \frac{|\widehat{\theta}-\overline{\theta}|}{|\overline{\theta}|}
\end{equation*}
as well as the root mean square error (RMSE) with respect to the original surface i.e.
\begin{equation*}
  \text{RMSE}(\widehat{\theta})=\sqrt{\sum_{i=1}^n 
    \sum_{j=1}^m(\widetilde{F}(\widehat{\theta})_{ij}-\sigma^{MKT}_{BS}(T_i,k_j))^2}.
\end{equation*}
Therefore, on one hand a measure of good calibration is a small RMSE. On the
other hand, a measure of parameter sensitivity on a given model is the
combined result of RMSE and parameter relative error. For this set of tests,
we again restrict ourselves to the image-based approach for learning the price
(implied volatility) function in the model.

\begin{figure}[H]
\hspace*{-0.5cm}\includegraphics[scale=0.5]{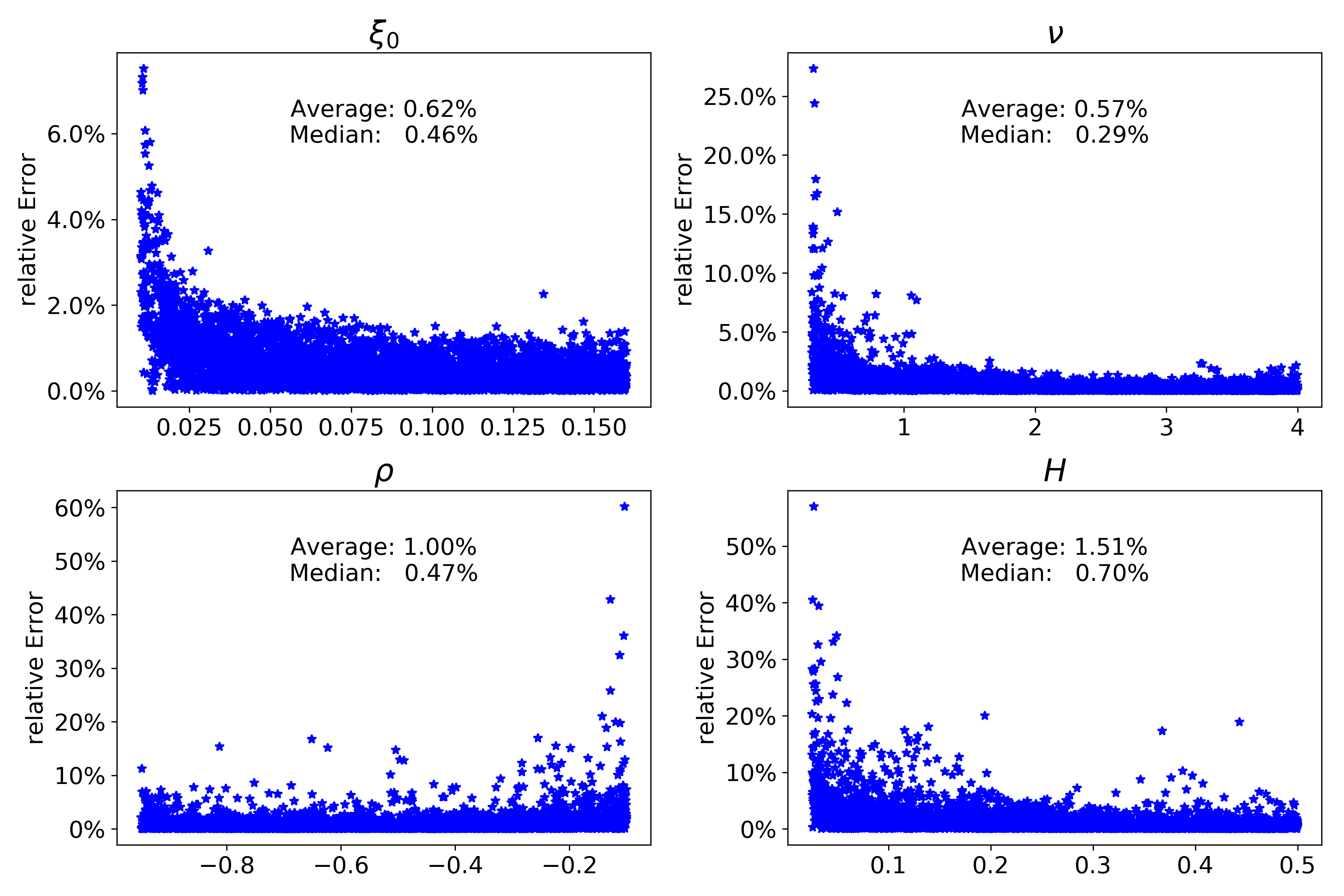}
\caption{Calibration relative error per parameter in the test set in the rough Bergomi model}
\label{fig:errors5}
\end{figure}
Figures \ref{fig:errors5} shows relative errors after calibration via
Levengerg-Marquardt in the rough Bergomi model. We observe that largest errors
are concentrated for small $H$ or small vol of vol $\nu$
situations. Naturally, the relative error is more sensitive around $0$ as
well. Once again, we emphasise that by understanding the error zones of the
pricing function $P$ (see Figure \ref{fig:rBergomi1}) along with parameter
relative errors in Figure \ref{fig:errors5}, we are able to asses its quality
and detect parameter configurations that might yield a lower performance of
the calibration process. 
\begin{figure}[H]
\hspace*{-2cm}\includegraphics[scale=0.4]{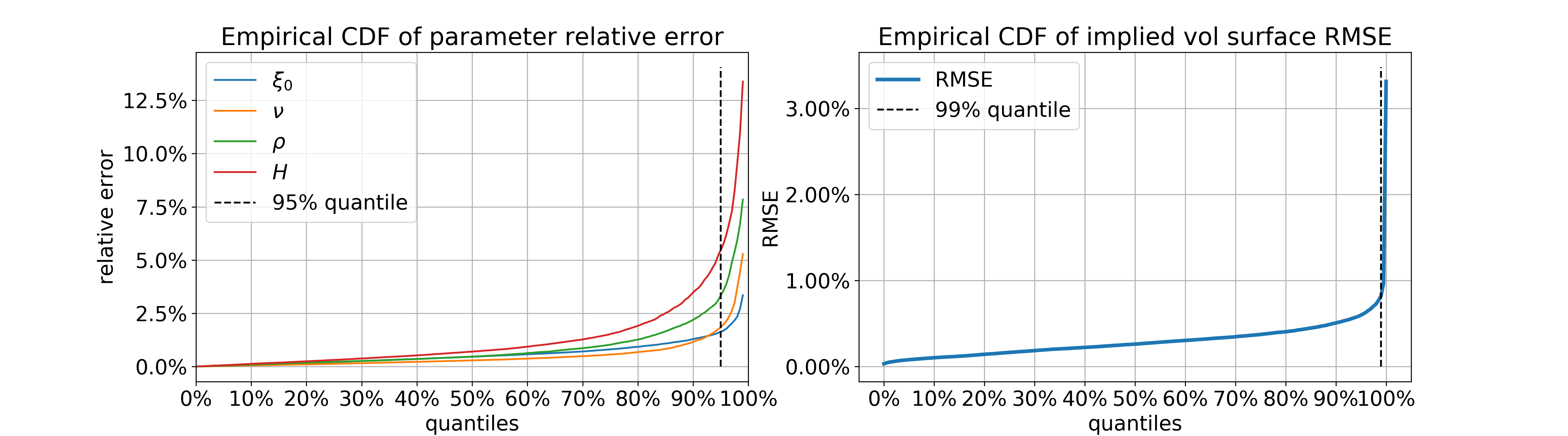}
\caption{Cumulative Distribution Function (CDF) of Rough Bergomi parameter
  relative errors (left) and RMSE (right) after Levengerg-Marquardt
  calibration across test set random parameter combinations.}
\label{fig:rBergomi4}
\end{figure}
To finalise our analysis, Figure  \ref{fig:rBergomi4} shows that the $99\%$ quantile of the RMSE is below $1\%$,
even though parameter relative errors might be higher (see \ref{fig:errors5}
as well), particularly when the parameters are close to $0$. Notably, the
maximum RMSE across the full surface (i.e. the 88 grid points) is below $4\%$, which suggests a surprisingly good
accuracy.

\subsection{A Bayes calibration experiment}
\label{sec:bayes-calibr-exper}

\begin{figure}
  \begin{center}
    \includegraphics[scale=0.5]{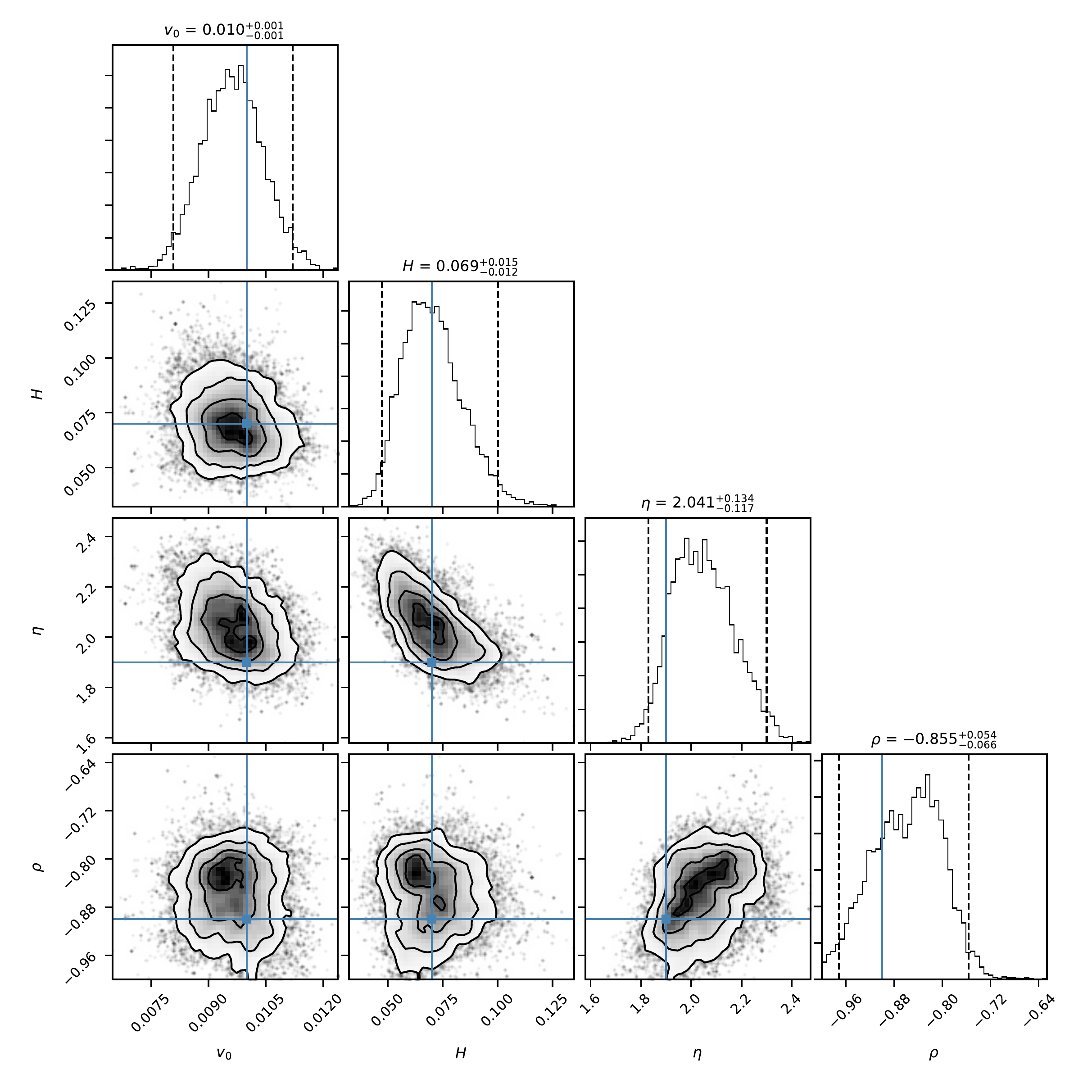}
  \end{center}
  \caption{Bayes calibration against synthetic implied volatility surface
    computed for model parameters $\theta^\dagger$. Solid vertical blue lines
    indicate true parameter values.}
  \label{fig:rb_bayes_calibration_synthetic}
\end{figure}

\begin{figure}
  \begin{center}
    \includegraphics[scale=0.5]{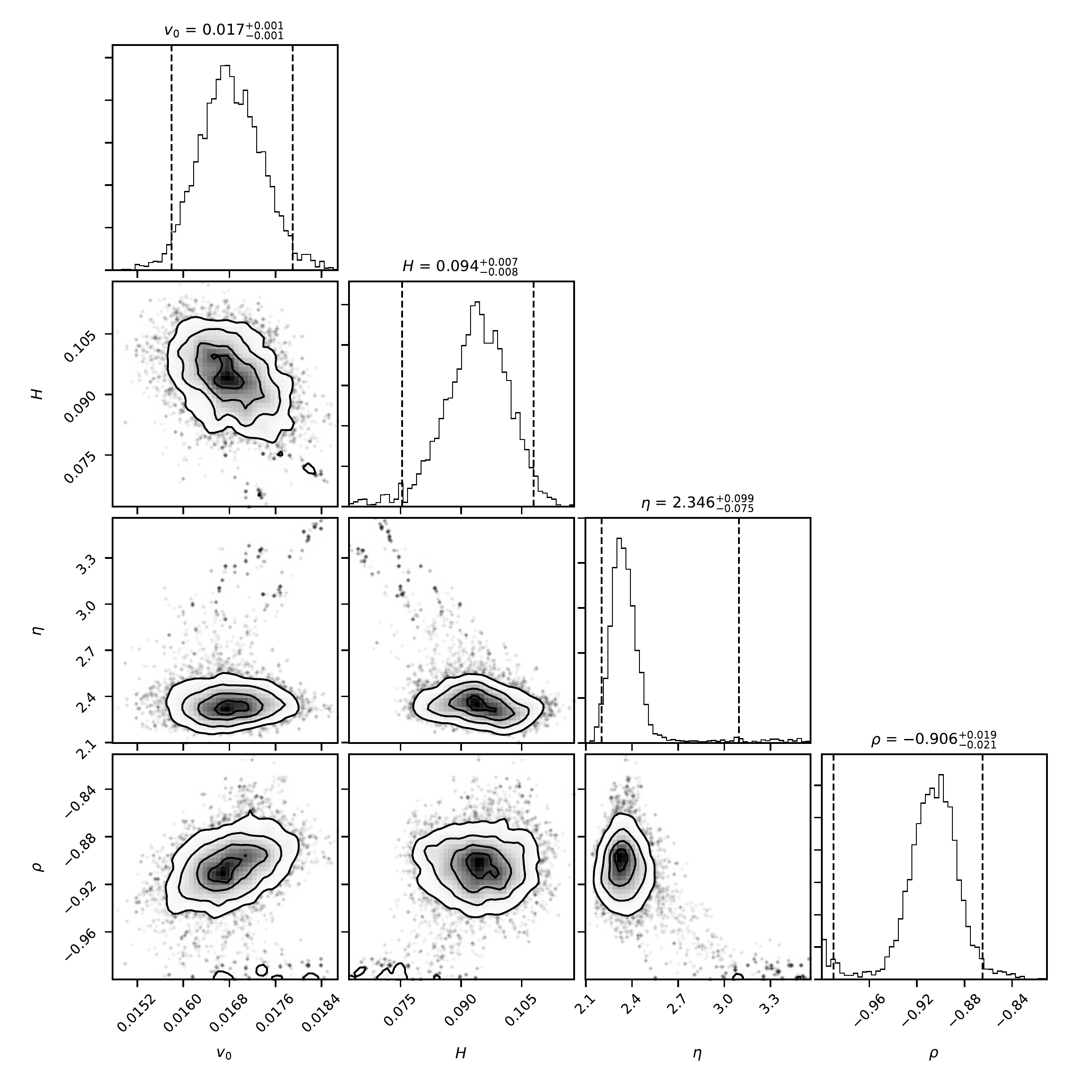}
  \end{center}
  \caption{Liquidity-weighted Bayes calibration against SPX market implied
    volatility surface from 19th May 2017. Liquidity proxies given by inverse
    bid-ask-spreads.}
  \label{fig:rb_bayes_calibration_marketdata}
\end{figure}

We next test the deep calibration procedure using the Bayesian point of view
sketched in Section~\ref{sec:bayes-analys-calibr}. Here, we use the pointwise
approach for learning the model implied volatility map.  We perform two
experiments. First, fixing $\theta = \theta^\dagger$, we generate a synthetic
implied volatility point cloud
\begin{equation*}
  \bm{y}_{\textrm{synth}}= \left[P\left(\theta^\dagger, T_1, k_1\right),
    \ldots, P\left(\theta^\dagger, T_N, k_N\right)\right] \in \mathbb{R}^N
\end{equation*}
using Monte Carlo simulation as in Section~\ref{sec:calibr-speed-accur}
above. Next, we perform a non-weighted Bayesian calibration against the
synthetic surface and collect the numerical results in Figure
\ref{fig:rb_bayes_calibration_synthetic}. 

More precisely, the figure shows histograms from the posterior distribution of
the one-dimensional marginal distribution of the (four-dimensional) parameter
$\theta$ in the rough Bergomi model, together with contour plots of all pairs
of two-dimensional marginal distributions based on kernel density estimates of
the joint densities. The titles of the histogram-windows report the empirical
medians together with the differences to the $2.5\%$ and $97.5\%$ quantiles,
respectively. The dashed lines in the histogram plots show those quantiles.

If the map $\widetilde{F}$ is sufficiently accurate for calibration, the
computed posterior should attribute a large probability mass around
$\theta^\dagger$. The results in Figure \ref{fig:rb_bayes_calibration_synthetic}
are quite striking in several ways: (1) From the univariate histograms on the
diagonal it is clear that the calibration routine has identified sensible
model parameter regions covering the true values. (2) Histograms are unimodal
and its peaks close or identical to the true parameters. (3) The isocontours
of the 2d Gaussian KDE in the off-diagonal pair plots for $(\eta, H)$ and
$(\eta, \rho)$ show exactly the behaviour expected from the reasoning in the
last section: Since increases or decreases in one of $\eta, H$ or $\rho$ can
be offset by adequate changes in the others with no impact on the calculated
IV, the Bayes posterior cannot discriminate between such parameter
configurations and places equal probability on both combinations. This can be
seen by the diagonal elliptic probability level sets.

In a second experiment, we want to check whether the inaccuracy of
$\widetilde{F}$ allows for a successful calibration against market data. To
this end, we perform a liquidity-weighted Bayesian regression against SPX
implied volatilities from 19th May 2017.  For bid and ask IVs $a_i > 0$ and
$b_i >0$ respectively, we proxy the IV of the mid price by
$m_i := \frac{a_i + b_i}{2}$.  With spread defined by
$s_i = a_i - b_i \geq 0$, all options with $s_i/m_i \geq 5\%$ are removed
because of too little liquidity. Weights are chosen to be
$w_i = \frac{m_i}{a_i - m_i} \geq 0$, effectively taking inverse bid-ask
spreads as a proxy for liquidity. Finally, $\sigma_i$ are proxied by a
fractional of the spread $s_i$. The numerical results in Figure
\ref{fig:rb_bayes_calibration_marketdata} further confirm the accuracy of
$\widetilde{F}$: (1) As can be seen on the univariate histograms on the diagonal,
the Bayes calibration has again identified sensible model parameter regions in
line with what is to expected. (2) Said histograms are again unimodal with
peaks at or close to values previously reported in the literature. (3) Quite
strikingly, at a first glance, the effect of the diagonal probability level
sets in the off-diagonal plots as documented in Figure
\ref{fig:rb_bayes_calibration_synthetic} cannot be confirmed here. However,
the scatter plots in the diagrams do reveal some remnants of that phenomenon.

\appendix
\section{A numerical experiment with the inverse map }
\label{sec:Inverse Map}

To motivate the main drawbacks of the inverse map approach of
Section~\ref{sec:hernandez}, we calibrate rough Bergomi model with it, i.e., we consider
the simple map
\begin{equation*}
  \Pi^{-1}(\Sigma^{\mathrm{rBergomi}}_{\mathrm{BS}})\to (\hat{\xi_0},\hat{\nu},\hat{\rho},\hat{H})
\end{equation*}
where $\Sigma^{\mathrm{rBergomi}}_{\mathrm{BS}} \in\mathbb{R}^{n\times m}$ is
a rBergomi implied volatility surface and $
(\hat{\xi_0},\hat{\nu},\hat{\rho},\hat{H})$ the optimal solution to the
corresponding calibration problem.

\begin{remark}
  For simplicity we consider the strikes and maturities to be fixed for all
  implied volatility surfaces.
\end{remark}

\textbf{Inverse Map Architecture}
\begin{itemize}
\item 1 convolutional layer with $16$ filters and $3\times3$ sliding window
\item MaxPooling layer with $2\times 2$ sliding window
\item 50 Neuron Feedforward Layer with \textit{Elu} activation function
\item Output layer with \textit{linear} activation function
\item Total number of parameters: $10.014$
\item Train Set: $34.000$ and Test Set: $6.000$
\item $(\xi_0,\nu,\rho,H)\in \mathcal{U}[0.01,0.16]\times
  \mathcal{U}[0.3,4.0]\times \mathcal{U}[-0.95,-0.1]\times
  \mathcal{U}[0.025,0.5]$ 
\item strikes=$\{0.5,0.6,0.7,0.8,0.9,1,1.1,1.2,1.3,1.4,1.5\}$
\item maturities=$\{0.1,0.3,0.6,0.9,1.2,1.5,1.8,2.0 \}$
\item Implied volatilities computed using Algorithm 3.5 in Horvath, Jacquier and
  Muguruza \cite{HJM17} with $60.000$ sample paths and the spot martingale
  condition i.e. $\mathbb{E}[S_t]=S_0,\;t\geq 0$ as control variate. 
\end{itemize}

\begin{figure}[H]
\includegraphics[scale=0.45]{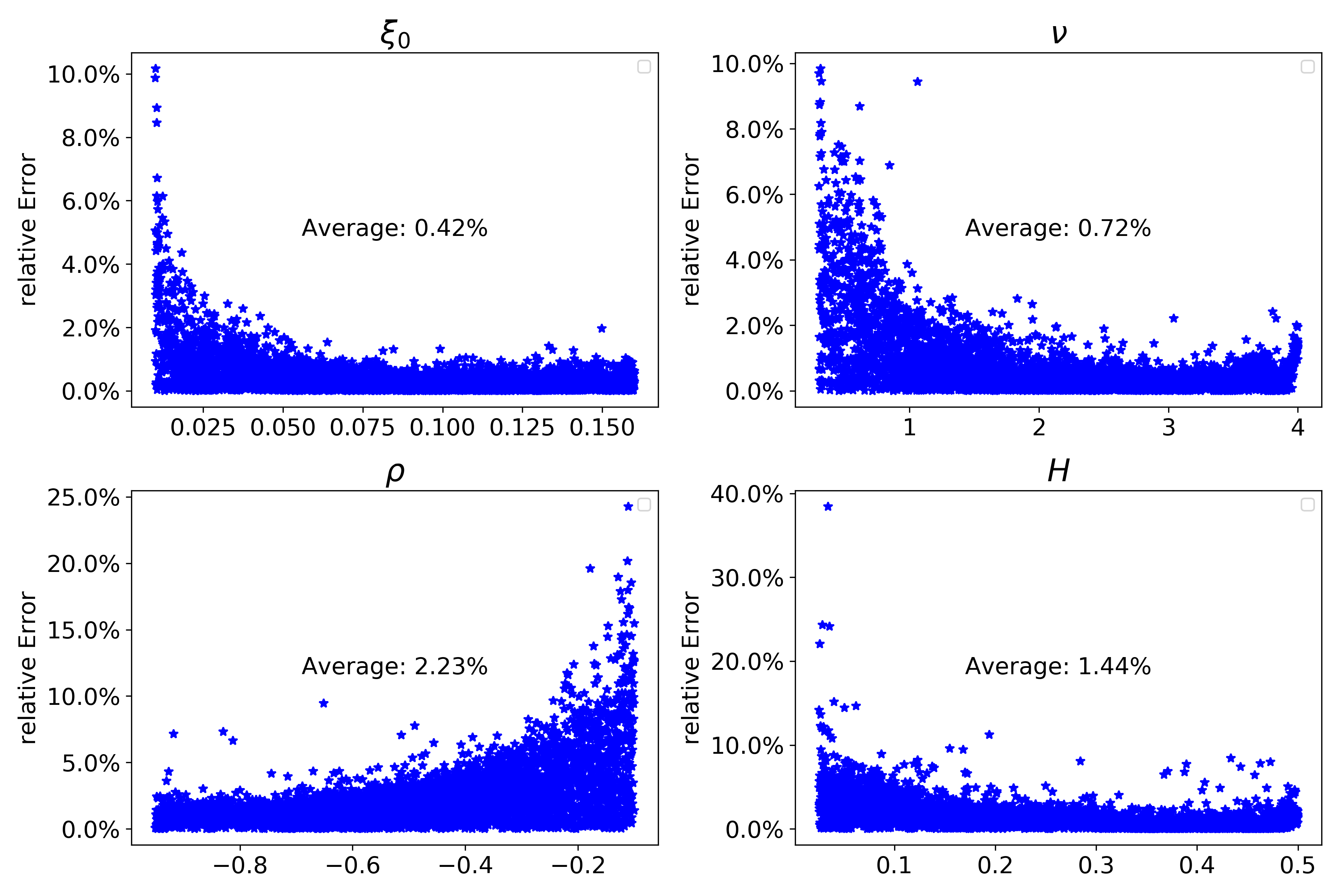}
\caption{Out of sample relative errors per parameter calibration}
\label{fig:errors1}
\end{figure}

Figure \ref{fig:errors1} shows that, indeed it is possible to approximate the
inverse map and very sharply calibrate model parameters with a relatively
small network. Convolutional networks make sense in this context, since a
implied volatility surface has many features both in the strike and maturity
direction that can be extracted, similar to image recognition problems. Notice
also that the biggest error come from parameter configurations where the Monte
Carlo input is more delicate i.e. very small $H$ or very small
volatility. Hence, the shape of the errors is intuitively natural and expected
beforehand.

\subsubsection*{Black-Box function and ``real'' out of sample performance}
\begin{figure}[H]
\hspace*{1.5cm}\includegraphics[scale=0.45]{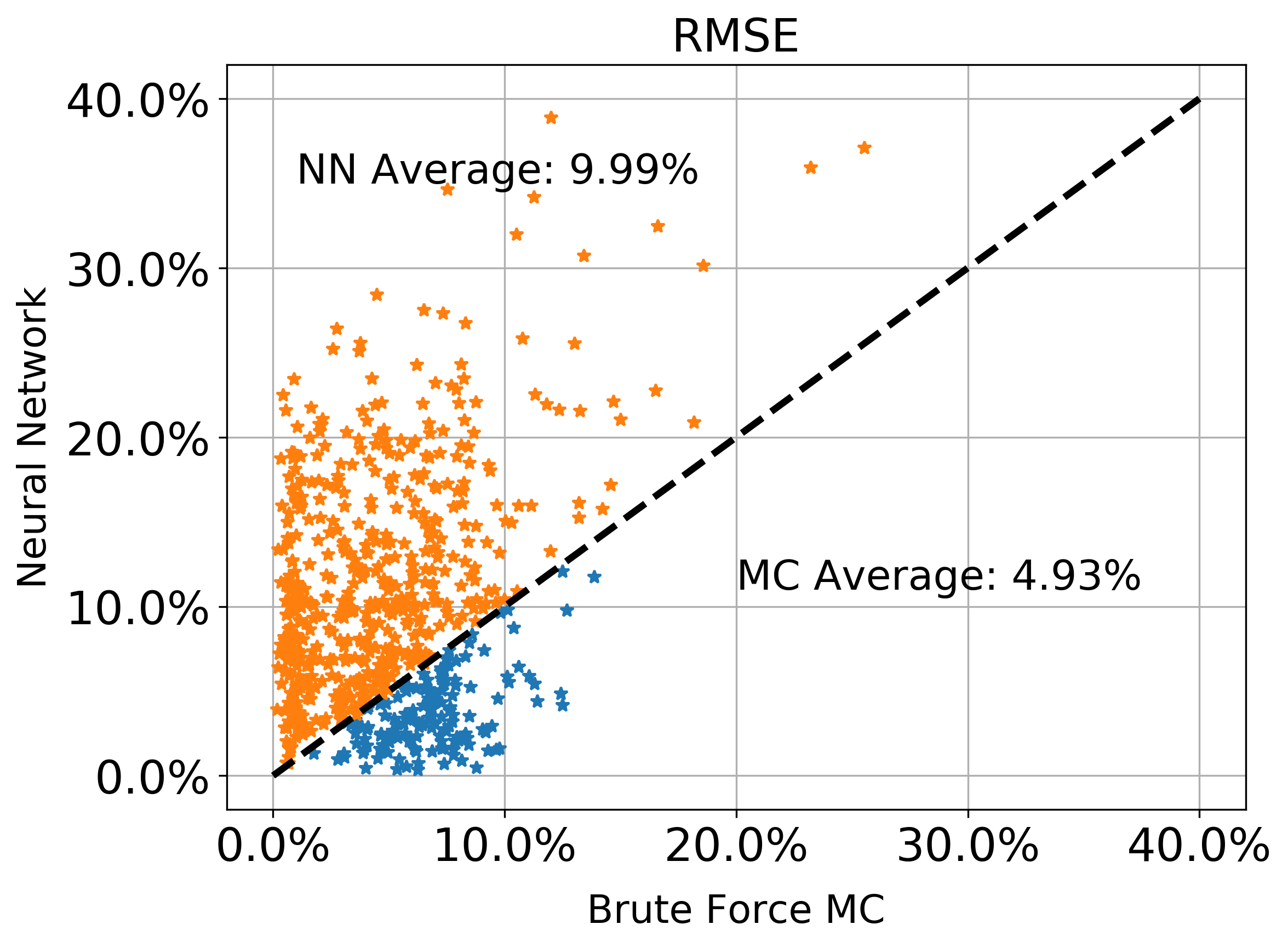}
\caption{ Stars represent the out of sample RMSE via neural network (NN) and
  brute force Monte Carlo (MC). Dashed black line represents the identity
  function.} 
\label{fig:errors3}
\end{figure}

Let us now consider ``real'' out of sample data, in the sense that it has not
been generated by the rough Bergomi model itself. We generate implied
volatility surfaces using the $2$ factor Bergomi given by

\begin{align*} dX_t&=-\frac{1}{2} V_t dt +\sqrt{V_t} dW_t \\
V_t&=\xi_0(t)\mathcal{E}\left(\nu \left((1-\theta)\int_0^t \exp(-\kappa_X(t-s))dZ_s+\theta \int_0^t \exp(-\kappa_Y(t-s))dY_s\right)\right)
\end{align*}

where $W$, $Y$ and $Z$ are correlated standard Brownian motions. We feed smiles from this model into our neural network to
obtain the corresponding optimal rough Bergomi parameters. We then compare
these values with a direct Monte Carlo calibration via Levenberg-Marquardt
\cite{Levenberg,Marquardt} algorithm. Figure \ref{fig:errors3} shows that the
neural network does not generalize properly out of sample and concludes that
the brute force MC method clearly beats the Inverse map approach. The results
by Hernandez \cite{Hernandez} also support this conclusion, since his out of
sample performance (based on different historical period) is reasonably worse
than the in-sample one. However, we must emphasize that when the neural
network is exposed to familiar situations i.e. surfaces close to the ones
generated by the rBergomi model it may work better than the MC approach (see
points below the dashed red line in Figure \ref{fig:errors3}). This is likely
due to delicate parameter configurations i.e. very low variance, where MC
suffers to obtain accurate estimates whereas the network does not struggle
that much.
The \emph{one-step} approach does not generalise the
problem to all possible settings, since by
design is not possible to train $\Pi^{-1}$ on all possible (arbitrage-free)
market scenarios. Moreover, there is a
lack of understanding in the highly non-trivial function $\Pi^{-1}$, hence from a
risk-managing perspective is more difficult to justify the use of this inverse
approach than of the direct approach. 

\newpage
\section{Illusration of model parameters \& the pricing engine in the rBergomi model}

\begin{figure}
\begin{center}
\includegraphics[scale=0.25]{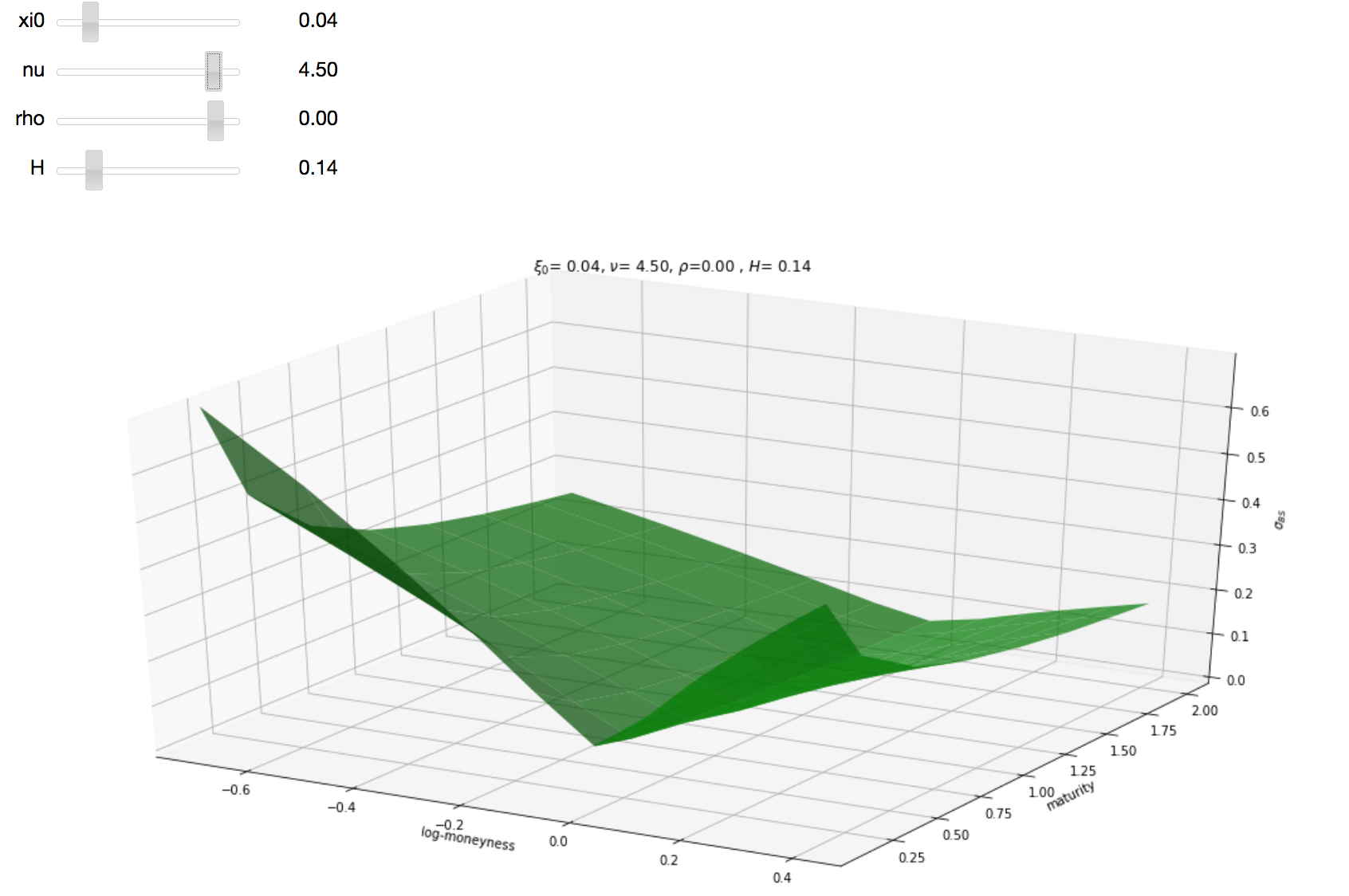}\includegraphics[scale=0.25]{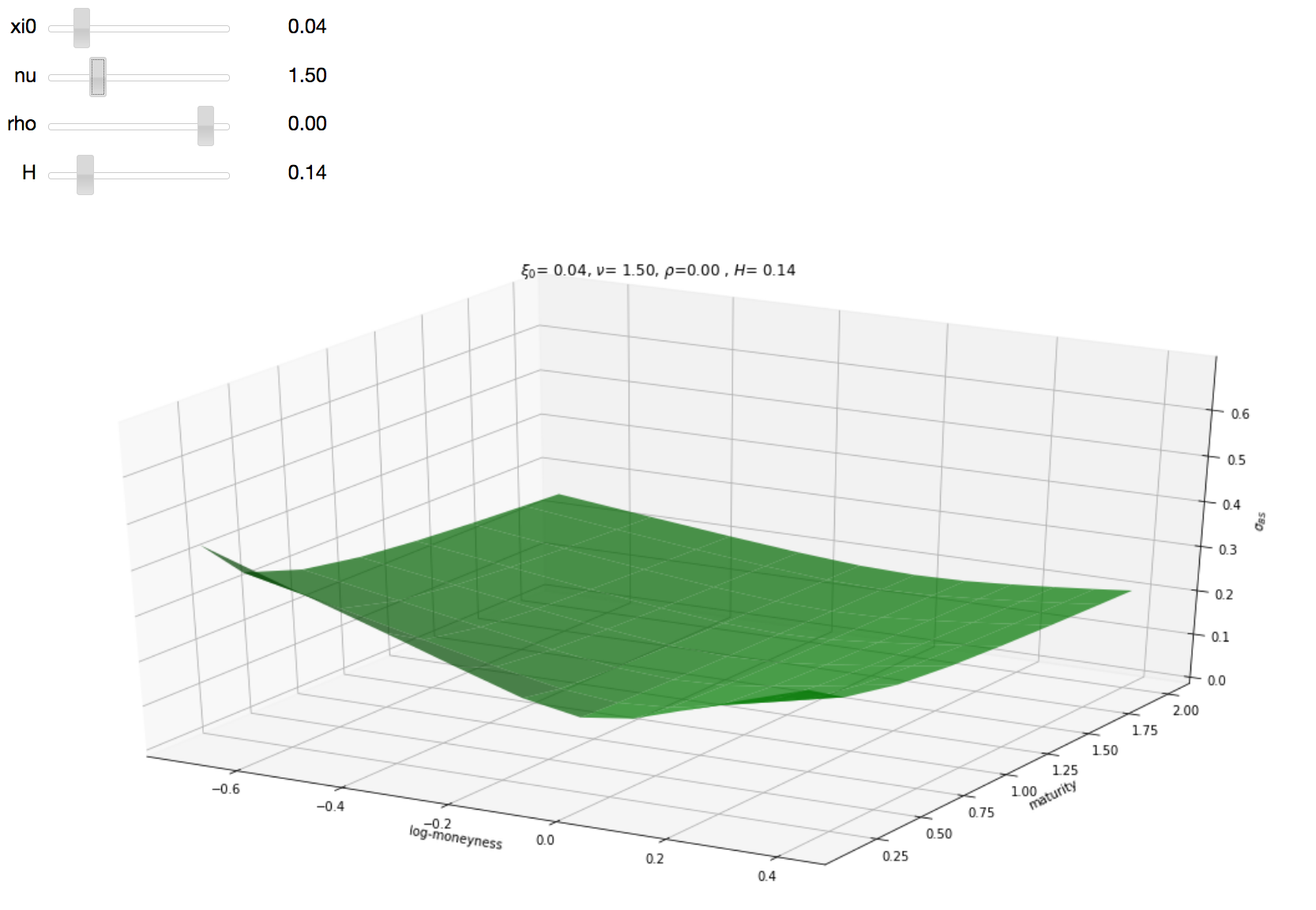}\\
\includegraphics[scale=0.25]{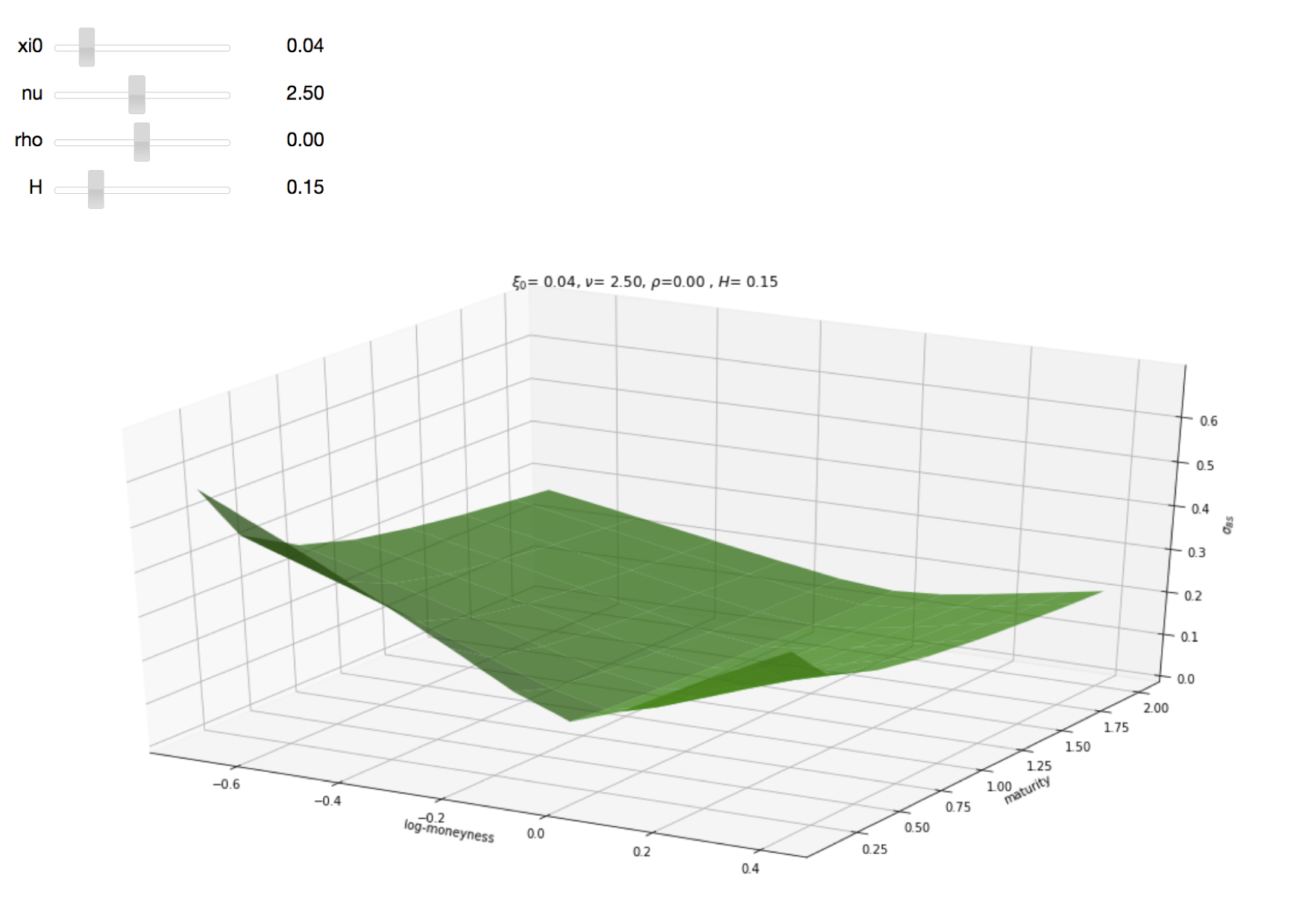}
\includegraphics[scale=0.25]{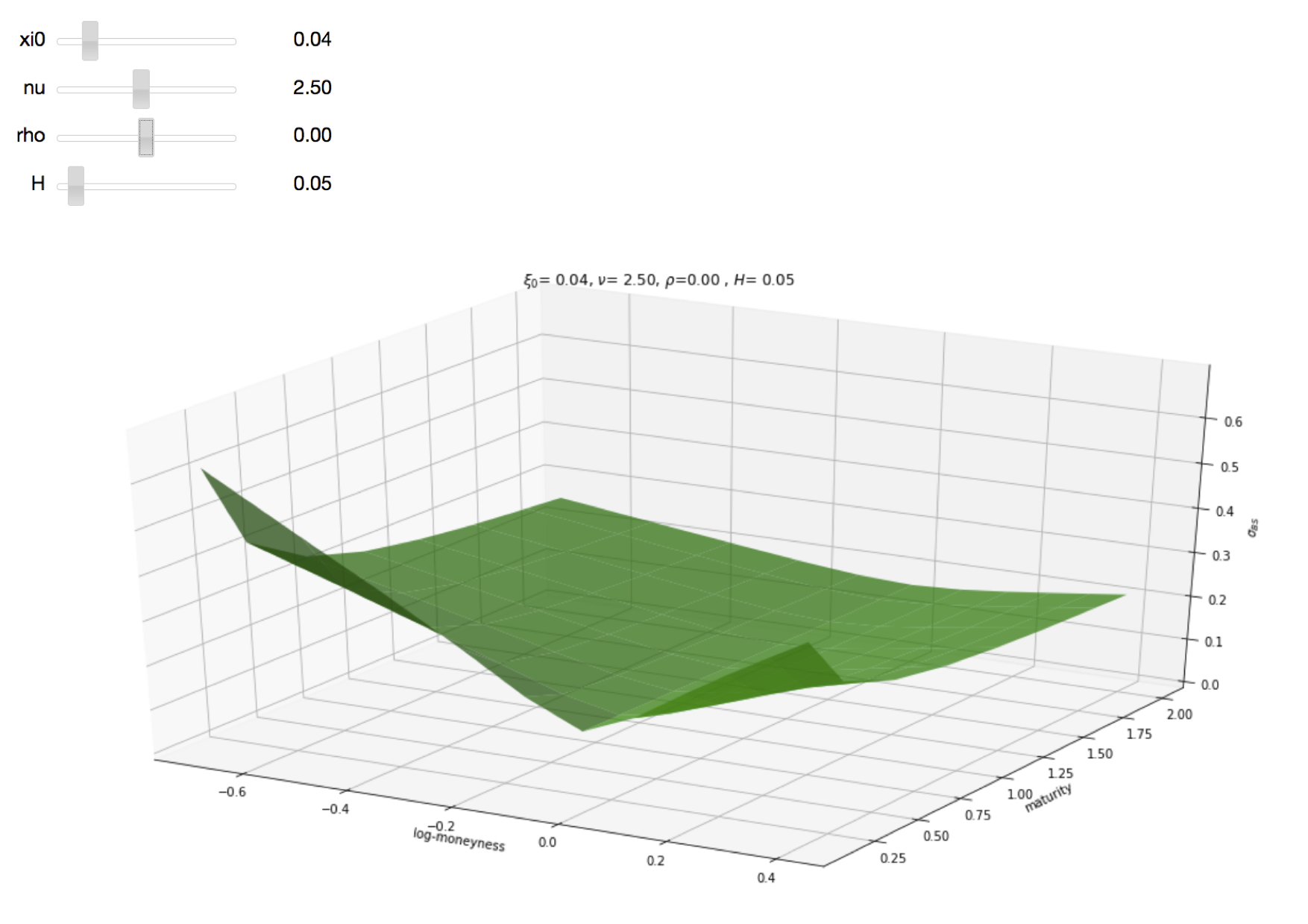}
\end{center}
\end{figure}
We showcase here the influence of the model parameters in the rough Bergomi model on the shape of the implied volatility surface using the hands-on pricing engine we generated via the DNN of \textbf{step (i)} for the rough Bergomi model. Our findings are in line with asymptotic results presented in \cite{BFGHS} and with \cite{MP18} for the role of the model parameters.
\vspace*{0.3cm}\\
The model parameters $\nu, \rho$ and $H$ correspond to the \emph{smile }($\nu$), \emph{skew }($\rho$) and the \emph{explosion ($H$) parameters} of the surface, while $\xi_0$ is the one-point approximation of the forward variance.\\
The images illustrate that the parameters $\nu$ and $H$ influence the slope of the smile, and an explosive behaviour for short maturities can be achieved (without calibrating slice by slice) with a single surface if the Hurst parameter is $H<<\frac{1}{2}$. 
And finally, as usual in stochastic volatility models, the parameter $\rho$ introduces skewness in the surface as illustrated below.

\begin{figure}
\begin{center}
\includegraphics[scale=0.25]{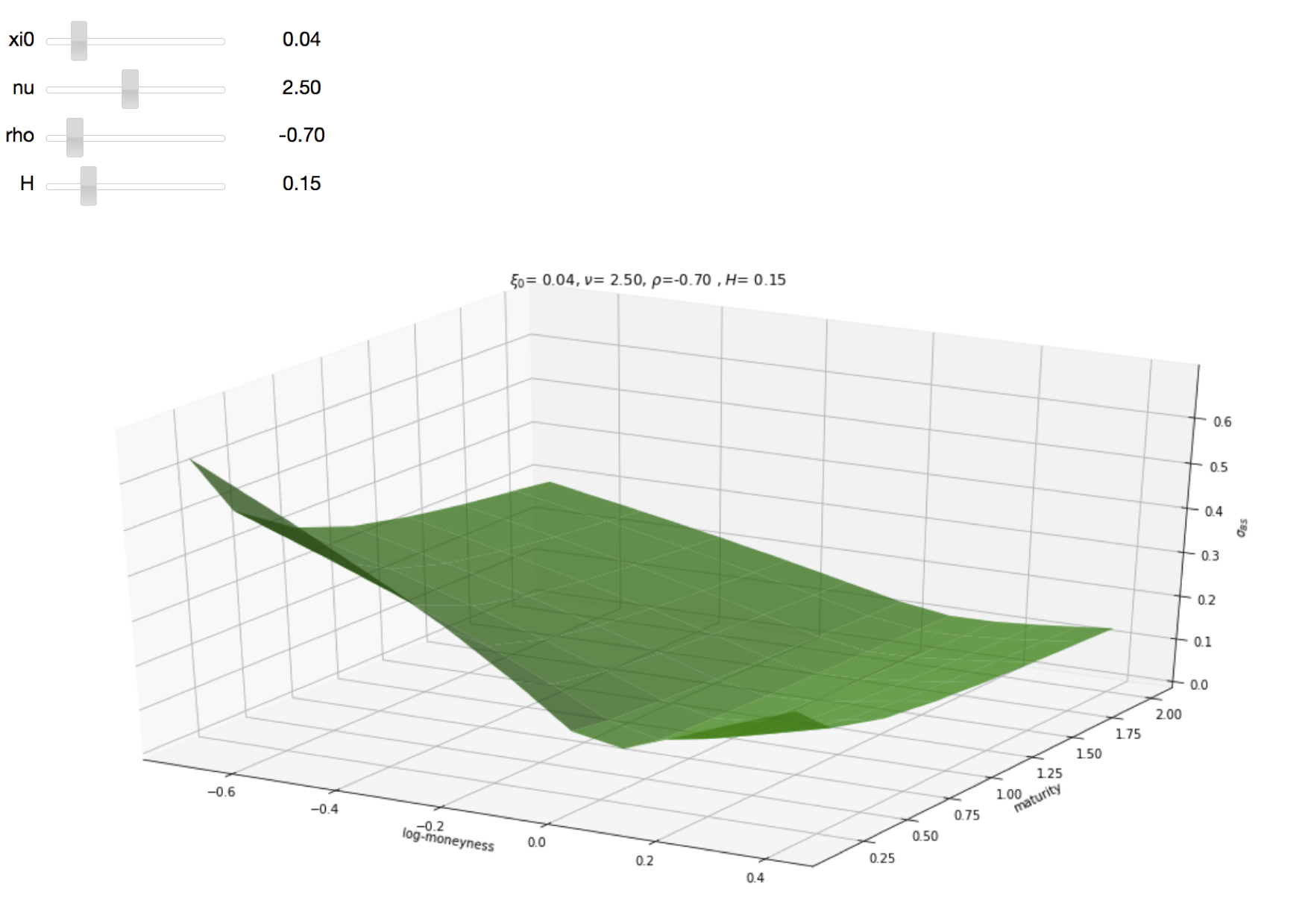}
\includegraphics[scale=0.25]{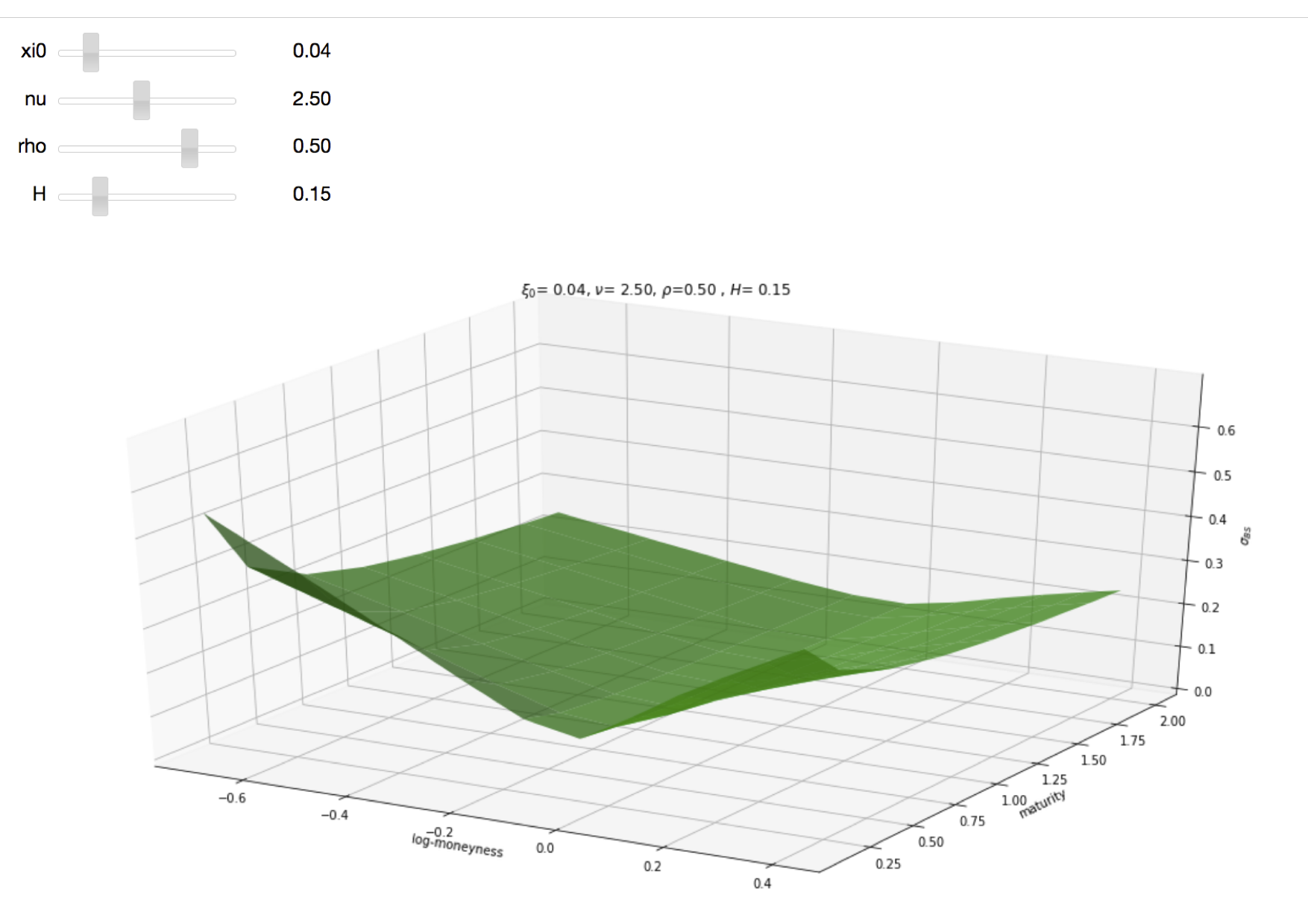}
\end{center}
\end{figure}
\newpage


\begin{thebibliography}{99}

\bibitem{AlosLeon}E. Al\`os, J. Le\'on and J. Vives. 
On the short-time behavior of the implied volatility for jump-diffusion models with stochastic volatility. 
\textit{Finance and Stochastics}, {\tt 11}(4), 571-589, 2007.

\bibitem{AKS19}A. Antonov, M. Konikov, M. Spector.
Modern SABR Analytics: Formulas and Insights for Quants, Former Physicists and Mathematicians.
\textit{SpringerBriefs in Quantitative Finance}, 2019.

\bibitem{ACS99}M. Avellaneda, A. Carelli, F. Stella.
Following the Bayes path to option pricing.
\textit{Journal of Computational Intelligence in Finance}, {\tt 8}4, 1999.


\bibitem{BFGMS17}C.~Bayer, P.~Friz, P.~Gassiat, J.~Martin and B.~Stemper.
A regularity structure for rough volatility. 
\href{https://arxiv.org/abs/1710.07481}{arXiv:1710.07481}, 2017.

\bibitem{BFG15}C.~Bayer, P.~Friz  and J.~Gatheral.
Pricing under rough volatility. 
\textit{Quantitative Finance}, {\tt 16}(6): 1-18, 2016.

\bibitem{BFGHS}C.~Bayer, P. Friz, A. Gulisashvili, B. Horvath and B. Stemper.
Short-time near the money skew in rough fractional stochastic volatility models.
\href{https://arxiv.org/abs/1703.05132}{arXiv:1703.05132}, 2017.


\bibitem{BS18}C.~Bayer and B.~Stemper. Deep calibration of rough stochastic volatility models. \textit{Preprint}, \href{https://arxiv.org/pdf/1810.03399.pdf}{arXiv:1810.03399}

\bibitem{BLP16} M.~Bennedsen, A.~Lunde and M.S.~Pakkanen. 
Hybrid scheme for Brownian semistationary processes. 
\textit{Finance and Stochastics}, {\tt 21}(4): 931-965, 2017.

\bibitem{BergomiBook}L.~Bergomi. Stochastic Volatility Modeling.
Chapman \& Hall/CRC financial mathematical series.  \textit{Chapman \& Hall/CRC}, 2015.



\bibitem{OsterleeI} B.~Chen, C.~W. ~Oosterlee and H.~Van~Der~Weide. Efficient unbiased simulation scheme for the SABR stochastic volatility model, 2011.


\bibitem{Cont10}R. Cont.
Model Calibration.
\textit{Encyclopedia of Quantitative Finance},
{\tt 5}, 2010. \href{https://doi.org/10.1002/9780470061602.eqf08002
}{DOI:10.1002/9780470061602.eqf08002}

\bibitem{CulcinDas17}R. Culkin and S. R. Das 
Machine Learning in Finance: The Case of Deep Learning for Option Pricing. 
\textit{Journal of Investment Management}, \href{https:srdas.github.io/Papers/BlackScholesNN.pdf}{github:BlackScholesNN} 2017.


\bibitem{MadanSchoutens} J.~De Spiegeleer, D.~Madan, S.~Reyners and W.~Schoutens. Machine learning for quantitative finance:
Fast derivative pricing, hedging and fitting. \href{https://papers.ssrn.com/sol3/papers.cfm?abstract_id=3191050}{SSRN:3191050}, 2018.

\bibitem{HestonConvolutional} G.~Dimitroff, D.~R\"oder and C.~P. Fries. Volatility model calibration with convolutional neural networks. \textit{Preprint}, \href{https://papers.ssrn.com/sol3/papers.cfm?abstract_id=3252432}{SSRN:3252432}, 2018.

\bibitem{DepthNN}R.~Eldan and O.~Shamir. The power of depth for feedforward neural neworks. \textit{JMLR: Workshop and Conference Proceedings} {\tt Vol 49:1-34}, 2016.



\bibitem{ER18}O. El Euch and M. Rosenbaum.
Perfect hedging in rough Heston models, to appear in
\textit{The Annals of Applied Probability}, 2018.

\bibitem{FZ17} M.~Forde, H.~Zhang. Asymptotics for Rough Stochastic Volatility

\bibitem{StatisticalLearning}J.~Friedman, R.~Tibshiran and  T.~Hastie. The Elements of Statistical Learning. \textit{Springer New York Inc}, 2001.

\bibitem{FG18} R.~Ferguson and A. D.~Green. Deeply learning derivatives.
  Preprint \href{https://arxiv.org/abs/1809.02233}{arXiv:1809.02233}, 2018.

\bibitem{FHLG13} D. Foreman-Mackey, D. W. Hogg, D. Lang, J. Goodman.
  emcee: the MCMC hammer, \emph{Publications of the Astronomical Society of
    the Pacific}, 125(925), 306, 2013.

\bibitem{For16} D. Foreman-Mackey, corner.py: Scatterplot matrices in Python,
  \emph{The Journal of Open Source Software} 24,
  \url{http://dx.doi.org/10.5281/zenodo.45906}, 2016.

\bibitem{Fukasawa}M. Fukasawa.
Asymptotic analysis for stochastic volatility: martingale expansion.
\textit{Finance and Stochastics}, {\tt 15}: 635-654, 2011.

\bibitem{Gat11} J.~Gatheral. The volatility surface: a practitioner's guide,
\textit{Wiley}, 2011.


\bibitem{GJR14}J.~Gatheral, T.~Jaisson  and M.~Rosenbaum.
Volatility is rough.
\textit{Quantitative Finance}, {\tt 18}(6): 933-949, 2018.


\bibitem{Glau19}K. Glau, D. Kressner, and F. Statti.
Low-rank tensor approximation for Chebyshev interpolation in parametric option pricing.
\href{https://arxiv.org/pdf/1902.04367.pdf}{arXiv:1902.04367}, 2019.

\bibitem{Green15}A. Green. XVA: Credit, Funding and Capital Valuation Adjustments. \textit{Wiley}, 2015.

%



\bibitem{Hagan1}P. Hagan, D. Kumar, A. Lesniewski, and D. Woodward. Managing smile risk. \textit{Wilmott Magazine}, {\tt September issue: 84-108}, 2002.



\bibitem{HJW17}J. Han, A. Jentzen, E. Weinan.
Overcoming the curse of dimensionality: Solving high-dimensional partial differential equations using deep learning.
\textit{PNAS}.{\tt 115}(34) 8505-8510, August 2018.


\bibitem{Hes93}S.L.~Heston. A closed-form solution for options with stochastic
  volatility with applications to bond and currency options, \textit{The Review of Financial Studies} 6(2):327--343, 1993.

\bibitem{Hernandez}A.~Hernandez. Model calibration with neural networks. \textit{Risk}, 2017.    


\bibitem{UniversalApprox}K. Hornik, M. Stinchcombe, and H. White. 
Multilayer feedforward networks are universal approximators. 
\textit{Neural Networks}, {\tt 2}(5):359-366, 1989.


\bibitem{UniversalApproxDerivatives}K.~Hornik. M.~Stinchcombe and H.~White. Universal approximation of an unknown mapping and its derivatives using multilayer feedforward networks. \textit{Neural Networks}.
Vol. 3:11, 1990.



\bibitem{PHL17}P. Henry-Labord\`{e}re.
Deep primal-dual algorithm for BSDEs: applications of machine learning to CVA and IM. 
\href{https://ssrn.com/abstract=3071506}{SSRN:3071506}

\bibitem{HJM17}B.~Horvath, A.~Jacquier and A.~Muguruza.
Functional central limit theorems for rough volatility.
\href{https://arxiv.org/abs/1711.03078}{arXiv:1711.03078}, 2017.

\bibitem{HMM19}B.~Horvath,  A.~Muguruza and T.~Mehdi.
Deep learning volatility.
Available at SSRN 3322085, 2019.

\bibitem{HR18}B. Horvath, O. Reichmann.
Dirichlet Forms and Finite Element Methods for the SABR Model.
\textit{SIAM Journal on Financial Mathematics}, {\tt p. 716-754}(2), May 2018.


\bibitem{HW90}J. Hull and A. White. 
Pricing interest rate derivatives securities. 
\textit{The Review of Financial Studies}){\tt(3)}: 573-592, 1990

\bibitem{Hutchison94}
J. M. Hutchinson, A. W. Lo and T. Poggio.
A Nonparametric Approach to Pricing and Hedging Derivative Securities Via Learning Networks.
\textit{The Journal of Finance}, {\tt 49}(3) 851-889. \textit{Papers and Proceedings Fifty-Fourth Annual Meeting of the American Finance Association, Boston, Massachusetts}, 1994.

\bibitem{Itkin14} A. Itkin.
To sigmoid-based functional description of the volatility smile.
\textit{Preprint}, \href{https://arxiv.org/pdf/1407.0256.pdf}{arXiv:1407.0256}, 2014. 


\bibitem{IS15}S. Ioffe and C. Szegedy.
Batch normalisation: Accelerating deep network training by reducing internal covariate shift. \textit{Preprint}, \href{https://arxiv.org/abs/1502.03167}{arXiv:1502.03167}, 2015.





\bibitem{KBAdam} D.P.~Kingman and J.~Ba, Adam: A Method for Stochastic Optimization. \textit{Conference paper}, 3rd International Conference for Learning Representations, 2015.





\bibitem{OsterleeII} A.~Leitao~Rodriguez, L.A.~Grzelak and C.W.~Oosterlee. On an efficient multiple time step Monte Carlo simulation of the SABR model. \textit{Quantitative Finance}, {\tt 17}(10), pp.1549-1565, 2017.

\bibitem{Levenberg} K.~Levenberg. A Method for the Solution of Certain Non-Linear Problems in Least Squares. \textit{Quarterly of Applied Mathematics}. 2: pp. 164-168, 1944.
\bibitem{Marquardt} D.~Marquardt. An Algorithm for Least-Squares Estimation of Nonlinear Parameters. \textit{SIAM Journal on Applied Mathematics}. 11 (2): pp. 431-441,1963.`

\bibitem{LiuGrezakOosterlee} S. Liu, A. Borovykh, L. A. Grzelak, C. W. Oosterlee.
A neural network-based framework for financial model calibration.
Preprint, \href{https://arxiv.org/abs/1904.10523}{arXiv:1904.10523}, 2019.


\bibitem{McGhee}W.~A.~McGhee. An artificial neural network representation of the SABR stochastic volatility model. \textit{Preprint}, \href{https://ssrn.com/abstract=3288882}{SSRN:3288882}, 2018.


\bibitem{MP18} R.~McCrickerd, M.~Pakkanen, Turbocharging Monte Carlo pricing for the rough Bergomi model, \textit{Quantitative Finance} 18(11):1877-1886, 2018.



\bibitem{RGLO17}A. Leitao Rodriguez, A. Grzelak Lech, Cornelis W. Oosterlee.
On a one time-step Monte Carlo simulation approach of the SABR model : Application to European options.
\textit{Applied Mathematics and Computation}, {\tt 293} p. 461-479, 2017,

\bibitem{SSSz18} M. Sabate Vidales, D. Siska, L. Szpruch
Unbiased deep solvers for parametric PDEs
\href{https://arxiv.org/abs/1810.05094}{arXiv:1810.05094}, 2018.


\bibitem{SS18}J. Sirignano and K. Spiliopoulos.
DGM: A deep learning algorithm for solving partial differential equations.
\textit{Journal of Computational Physics}
{\tt 375}(15) 1339-1364, December 2018


\bibitem{SW2011} L.~Setayeshgar, and H.~Wang. Large deviations for a feed-forward network, \textit{Advances in Applied Probability}, 43: 2, pp. 545-571, 2011.

\bibitem{ShaClCo18}U. Shaham, A. Cloninger, and R. R. Coifman. 
Provable approximation properties for deep neural networks. 
\textit{Appl. Comput. Harmon. Anal.}, {\tt 44}(3): 537-557, 2018.

\bibitem{Stone}H. Stone. Calibrating rough volatility models: a convolutional neural network approach. \textit{Preprint}, \href{https://arxiv.org/pdf/1812.05315.pdf}{arXiv:1812.05315}, 2018.



\end{thebibliography}
\end{document}